%% file: si_main.tex
\documentclass[11pt]{article}

\usepackage[T1]{fontenc}
\usepackage[utf8]{inputenc}
\usepackage[english]{babel}
\usepackage{csquotes}
\usepackage[a4paper,margin=1in]{geometry}
\usepackage{longtable}
\usepackage{physics}
\usepackage{amsmath,amssymb}
\usepackage[version=4]{mhchem}
\usepackage{pdfpages}
\usepackage{graphicx}
\usepackage{caption}
\usepackage{booktabs}

\usepackage{multirow}
\usepackage{makecell}
\usepackage{authblk}

\usepackage{xcolor}
\usepackage[hidelinks]{hyperref}

\usepackage[
  backend=bibtex,  
  style=nature,
  citestyle=numeric-comp,
  autocite=superscript,
  sorting=none,          
  doi=false,
  url=false,
  isbn=false
]{biblatex}
\AtEveryBibitem{%
  \clearfield{month}%
  \clearfield{day}%
  \clearfield{endmonth}%
  \clearfield{endday}%
  \clearfield{urlmonth}%
  \clearfield{urlday}%
}

\let\cite\autocite

\newcommand{\rev}[1]{#1}

\newif\ifhighlight

\newcommand{\highlight}{\highlighttrue}
\highlight
\newcommand{\editor}[2]{%
  \expandafter\newcommand\csname #1note\endcsname[1]{%
    \textcolor{#2}{(\textbf{#1note:} \textsc{##1})}}%
  \expandafter\newcommand\csname #1\endcsname[1]{%
    \ifhighlight\textcolor{#2}{##1} \else ##1\fi}%
  \expandafter\newcommand\csname #1cancel\endcsname[1]{%
    \ifhighlight\textcolor{#2}{\sout{##1}}\fi}%
  \expandafter\newcommand\csname #1change\endcsname[2]{%
    \ifhighlight\textcolor{#2}{\sout{##1} ##2}\else ##2\fi}%
  \newenvironment{#1text}{\ifhighlight\color{#2}\fi}{\color{black}}
}

\editor{MC}{red}
\editor{MK}{orange}
\title{Supplementary Information for:\\[0.25em]Quantum-corrected NMR crystallography at scale}
\date{} 

\author[1]{Matthias Kellner$^{\square}$}
\author[2]{Ruben Rodriguez-Madrid$^{\square}$}
\author[2]{Jacob B. Holmes}
\author[1]{Victor Paul Principe}
\author[1]{Seio~Inoue}
\author[2]{Lyndon Emsley\thanks{Email: \href{mailto:lyndon.emsley@epfl.ch}{lyndon.emsley@epfl.ch}}}
\author[1]{Michele Ceriotti\thanks{Email: \href{mailto:michele.ceriotti@epfl.ch}{michele.ceriotti@epfl.ch}}}

\affil[1]{Laboratory of Computational Science and Modeling, Institut des Mat\'eriaux, \'Ecole Polytechnique F\'ed\'erale de Lausanne, 1015 Lausanne, Switzerland}
\affil[2]{Laboratory of Magnetic Resonance, Institut des Sciences et Ing\'enierie Chimiques, \'Ecole Polytechnique F\'ed\'erale de Lausanne, 1015 Lausanne, Switzerland}

\IfFileExists{preamble.tex}{\input{preamble}}{%
  \IfFileExists{preamble}{\input{preamble}}{}%
}

\begin{document}



# Quantum-corrected NMR crystallography at scale

Matthias Kellner,[1,+] Ruben Rodriguez-Madrid,[2,+] Jacob B. Holmes,[2] Victor Paul Principe,[1] Lyndon Emsley,[2,*] and Michele Ceriotti[1,*]

[1] Laboratory of Computational Science and Modeling, Institut des Matériaux, École Polytechnique Fédérale de Lausanne, 1015 Lausanne, Switzerland

[2] Laboratory of Magnetic Resonance, Institut des Sciences et Ingénierie Chimiques, École Polytechnique Fédérale de Lausanne, 1015 Lausanne, Switzerland

**ABSTRACT:** Structure determination by chemical-shift-driven NMR crystallography relies on comparing chemical shieldings measured in solid-state NMR experiments with simulations. However, computational cost limits the accuracy of shielding predictions, that usually rely on low-level electronic-structure approximations and neglect thermal and quantum mechanical nuclear motion, leading to large errors, especially for highly informative hydrogen-bonded protons. To address these limitations, we introduce a quantum-nuclei-corrected (QNC-NMR) approach. We generate inexpensively quantum ensembles using PET-MOLS, a novel machine-learning learning model of the interatomic potential transferable across molecular crystals. Using them as inputs to a chemical-shift model results in a two-fold improvement of the agreement with experiments for hydrogen-bonded protons, without the need for empirical corrections. The ability to sample structures consistent with the experimental conditions enables further refinement of the shielding model by finetuning it against measured shieldings. The favorable scaling with system size allows similar improvements for amorphous materials that are otherwise inaccessible to explicit DFT simulations.

## INTRODUCTION

Hydrogen bonding is a key structural motif that determines the packing, conformational preferences, and overall three-dimensional structure of soft and condensed organic matter.[1-4] However, determining the structure of hydrogen bonding motifs with high accuracy is challenging. This is particularly true since hydrogen atoms, with their low electron density, are practically transparent to X-rays.[5]

Solid-state nuclear magnetic resonance (NMR) chemical shifts, in combination with computational methods and diffraction, can be used to routinely elucidate the structure of molecular solids from powder samples.[6-9] The pronounced sensitivity of NMR shifts to hydrogen-bonding patterns, together with the applicability of solid-state NMR to amorphous systems, makes it particularly well suited to probe hydrogen-bonded structures in both crystalline and disordered condensed organic matter.[10]

Structure determination by chemical-shift driven NMR crystallography relies on comparison of experimentally measured chemical shifts with predictions from complementary *ab initio* calculations that link candidate structural motifs to measured shifts and, in favorable cases, enable full structure determination.[7-9,11,12] This approach has been successfully applied to resolve the structures of a range of complex materials including cements,[13,14] photovoltaic perovskites,[15,16] or powdered and amorphous active pharmaceutical ingredients, providing a robust basis for the analysis of amorphous drug formulations.[17-22]

Paradoxically, chemical shift computations are least reliable for precisely those labile protons and hydrogen-bonding environments where NMR measurements might be most valuable. In the literature, numerous cases have been reported where computed chemical shifts of labile protons significantly deviate from experimental values in both crystalline[23-25] and amorphous forms.[18,20] Such systematic mismatches directly impact the robustness of chemical shift-driven structure determination protocols.

Various factors contribute to this discrepancy. First, experimentally measured chemical shifts are ensemble averages over fast (sub-ns) vibrational and thermal motions, and theoretical predictions must be likewise formulated and computed as averages over ensembles of structures rather than only for the single minimum-energy structures that are typically used due to the computational cost of predicting and sampling the correct thermodynamic ensemble. Notably, the underlying interatomic potential used to generate ensembles of structures must permit bond breakage, which becomes essential in fractional protonation states and hydrogen bond rearrangements. Second, light nuclei – especially protons – are subject to further delocalization due to nuclear quantum effects. Third and finally, any practical approach must be scalable to large

systems containing thousands of atoms to enable the study of realistic complex structures.

Several existing strategies and previous work partially address these limitations. Early works, computing harmonic corrections employing classical forcefields, find significant improvement in the agreement between calculated and predicted values, but cannot correctly describe proton transfer and quantum delocalization.[26-28] *Ab initio* molecular dynamics, especially in their path integral formulation[29] commonly referred as path integral molecular dynamics (PIMD), as performed by Dračínsky, Hodgkinson and Bour in molecular organic solids, showed better agreement of predictions with experimental data.[30,31] This approach enables rigorous treatment of these effects but prevents studies of larger systems and is typically limited to GGA DFT functionals, which do not reliably yield accurate dissociation energetics.[32]

Advances in machine learning (ML) models have removed the computational cost barrier for predicting both shieldings and interatomic forces. Specifically, over the last decade, machine learning interatomic potentials (MLIPs) have emerged as an alternative to overcome the computational expenses of *ab initio* calculations. Those models can rapidly compute energies and atomic forces for both stable and distorted structures, and they can be trained on high-accuracy DFT calculations.[33] Fast and accurate ML models for chemical shielding predictions in molecular solids have also been trained on DFT computed shieldings,[34-38] due to the lack of large curated databases of experimentally measured chemical shifts. These ML models reduce the computational cost of shift calculations with respect to the equivalent DFT methods by several orders of magnitude while, for the ShiftML3 model, keeping accuracies of 0.53 ppm and 2.44 ppm for $^1$H and $^{13}$C, respectively against experimental benchmark sets.[37]

Combining both ML shielding models and MLIP for structural sampling allows speeding up chemical shielding computations of static structures and including the effects of finite temperature as well as quantum nuclear delocalization through PIMD simulations. This makes them orders of magnitude faster than *ab initio* quantum calculations and renders them accessible for both longer simulation times and larger systems. Engel *et al.* were the first to report chemical shift computations using jointly bespoke chemical shielding ML models and MLIPs, to describe the effect of finite-temperature and quantum nuclear fluctuations in NMR crystallography for molecular organic solids.[39] Similarly, Laurila *et al.* also showed that combining MLIP and ML-shielding models leads to improved $^{13}$C chemical shift predictions for the different $C_{60}$ isomers.[40] So far, the application of MLIPs to compute NMR observables has been largely limited to specific systems, as traditional MLIPs lack the transferability to general condensed organic matter. More recently, several "universal" MLIPs for either general materials or organic molecules were introduced. These models are commonly trained on large corpora of DFT reference calculations, intended to cover a broad range of compositional space. They yield satisfactory accuracy for general materials modeling problems within their training-domain.[41-45] The appeal of universal MLIPs lies in their transferability. Like classical all-atoms force fields[46] universal molecular MLIPs can predict interatomic potentials for almost any organic crystal, without losing much of the *ab initio*-quality bespoke MLIPs promise, nor the ability of treating reactive, bond-breaking events.[41-45]

In this work we introduce an approach that we dub quantum nuclei corrected NMR (QNC-NMR), to include a description of quantum nuclear effects on labile protons and their $^1$H chemical shifts in molecular organic solids. Figure 1 illustrates the workflow of the new approach. This is enabled by combining an ML model for chemical shift predictions (ShiftML3) with a newly developed transferable MLIP for molecular organic solids, PET-MOLS, to compute chemical shift averages from PIMD simulations of large-scale systems. We first test the method on a benchmark set of experimentally measured $^1$H shifts of organic crystals and on a few challenging case studies for crystalline structures, where we find that the QNC-NMR approach yields a factor two better agreement with experiment for hydrogen-bonded protons. We then apply the new protocol to the prediction of $^1$H chemical shifts of labile protons in large scale amorphous structural models of a glucagon-like peptide-1 receptor agonist (GLP-1RA).[20]

Our analysis relies heavily on the accurate computation of fractional protonation occupancies, which are sensitive to prediction errors of the potential energy, and we demonstrate that the uncertainty estimator of the MLIP is able to identify when the potential energy surface predicts highly uncertain protonation states with respect to the training reference DFT computations.

Finally, we introduce an "ensemble of latent features" (ELF) approach that allows us to reconcile the computational and experimental description of NMR measurements and to build a Δ-ShiftML model that achieves better accuracy by finetuning against experimental references.

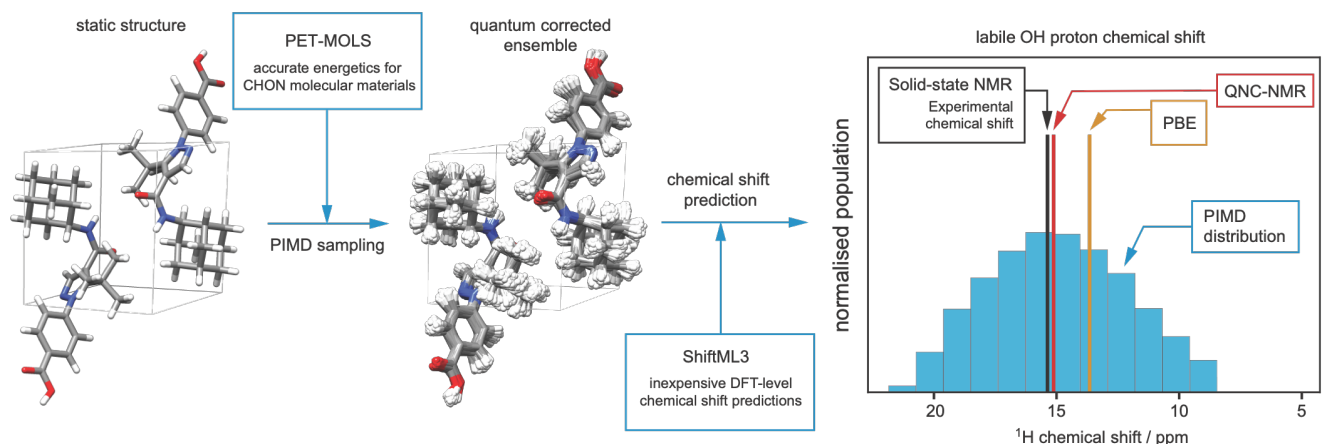

**Figure 1.** Schematic illustration of the steps for large scale QNC-NMR corrected shift predictions. First, selection of a candidate crystal structure (here Form 4 of AZD8329),[47] where DFT(PBE) chemical shift predictions for hydrogen-bonded protons (here shown by the orange line in the graph on the right for the OH proton) might typically have errors of ~1.5 ppm with respect to the experimentally determined chemical shifts. A PIMD sampling of the static form is run using the MLIP introduced here, dubbed PET-MOLS to generate an ensemble of structures. Chemical shifts are then predicted using the ShiftML3 model, which yields a distribution of chemical shifts (blue histogram). The average of the distribution, dubbed QNC-NMR approach (red line) is now in significantly better agreement with the experimental value.

## RESULTS

### A transferable potential for organic solids

To be able to efficiently generate ensembles of structures requires an interatomic potential that has the accuracy of DFT and the computational efficiency of an empirical forcefield, and which is transferable across a broad range of organic solids. To this end, we introduce PET-MOLS a ML potential trained on ~10,000 thermally sampled and relaxed configurations of diverse molecular crystals chosen from the Cambridge structural database (CSD),[48] that include H, C, N and O atoms.

The structures have been previously sampled to generate the ShiftML2 and ShiftML3 DFT shielding calculation training data.[48,37] We choose the PBE0+MBD DFT level of theory to generate training data for the ML potential, as PBE0 and hybrid DFT functionals correctly estimate the stability of fractional protonation states[32,49] and more generally yield unbiased bond length estimates in organic solids over commonly employed GGA functionals.[50] More details about ML model training and the reference computations are described in the Methods section and in SI. On hold-out test DFT data PET-MOLS achieves a test set accuracy of 3.3 meV/atom for energy and 62.4 meV/Å for forces, and provides excellent accuracy for modeling tasks beyond those we use here. A thorough characterization and validation of the model can be found in the SI.

### Chemical shift prediction for crystal structures

Figure 2 shows comparisons between three protocols for predicting (using ShiftML3[37]) $^1$H chemical shifts, benchmarking them on a set of 7 experimentally determined crystalline structures. The first protocol, dubbed Static-PBE, consists of running the predictions on the geometry relaxed structures using density functional theory (DFT) using the PBE density functional (see Section S1). This approach corresponds to the current standard in the field. The second one, dubbed Static-PBE0, follows the same approach as the first, but uses our new PET-MOLS for the geometry optimization. The last protocol is our QNC-NMR scheme, which consists of computing the averaged chemical shift predictions along a PIMD using PET-MOLS at constant temperature, set to that of the experimentally determined chemical shifts for each structure, and volume; commonly referred as NVT or a canonical ensemble (See detailed information in Methods).

Figure 2 shows that for the Static-PBE-relaxed structures, the overall root mean squared error (RMSE) between chemical shift predictions and experiment for all the $^1$H atoms in the set is 0.66 ppm, with a deviation of 1.63 ppm observed specifically for the set of hydrogen-bonded protons with experimentally measured chemical shifts greater than 9.6 ppm.

Using the Static-PBE0-relaxed geometries reduces the overall $^1$H error by about 14% (Figure 2b). PET-MOLS was trained on structures calculated at the PBE0 level, and as such these findings are consistent with Ramos *et al.*[50] who reported a 20–30% improvement in prediction accuracies of carbon and nitrogen GIPAW PBE computed shieldings from PBE0-optimized geometries, which are known to yield more-accurate and unbiased intramolecular bond lengths. In line with this, we also observe an improvement of about 28% for carbon shifts when going from Static-PBE to Static-PBE0 geometries (Figure S1).

Figure 2c shows how a significant further improvement is achieved by incorporating nuclear quantum corrections using the QNC-NMR protocol. Relative to the reference predictions for Static-PBE geometries, this reduces the errors by 25%, (0.66 ppm to 0.50 ppm) for hydrogen and 32% for carbons (2.38 ppm to 1.81 ppm) for all atoms in the experimental benchmarks. As expected, the improvement is particularly

remarkable for the H-bonded sites, for which the prediction error is more than halved (1.63 to 0.75 ppm). On the other hand, accuracies for non-hydrogen-bonded protons remain the same as for the Static-PBE0 geometries.

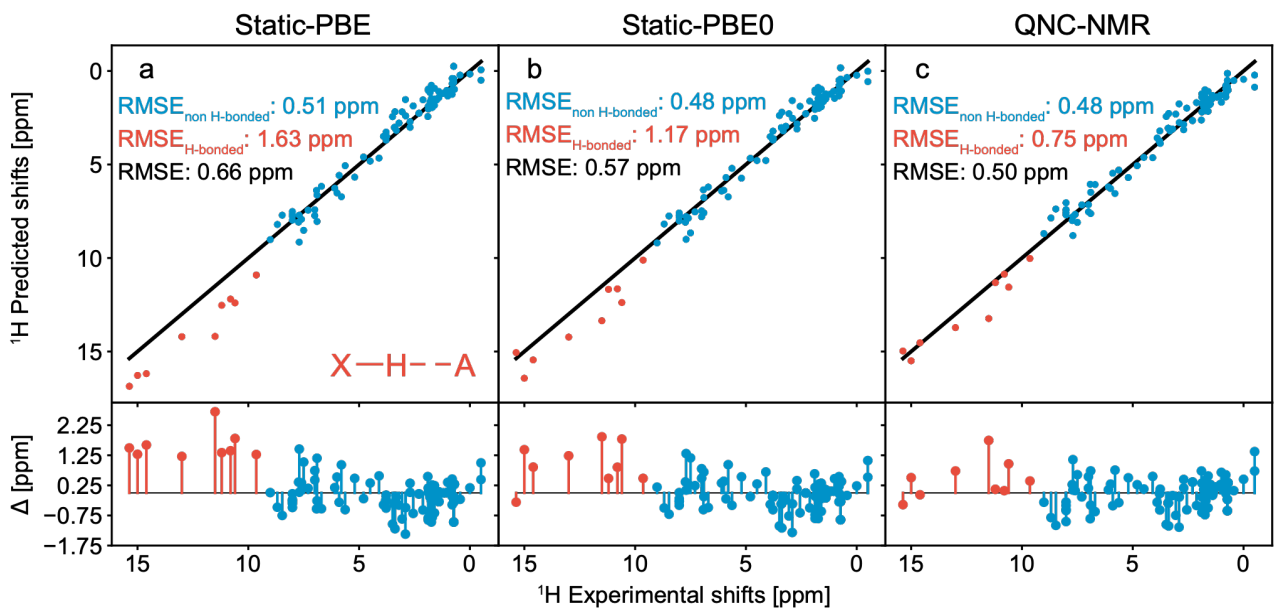

**Figure 2.** Comparison of experimental and predicted $^1$H chemical shifts computed with ShiftML3, using Static-PBE optimized geometries (a), Static-PBE0 optimized geometries (b), and the QNC-NMR approach (c). The black diagonal line in each panel corresponds to a perfect correlation. Red points mark hydrogen-bonded protons in the experimental benchmark set whose experimental chemical shift is greater than 9.6 ppm. The accuracy of the predictions is evaluated with the root.-mean square error (RMSE) of all $^1$H atoms (in black), the non H-bonded protons (RMSE$_{\text{non H-bonded}}$, in blue), and the H-bonded protons (RMSE$_{\text{H-bonded}}$, in red). On the lower panels, $\Delta$ is the difference between the predicted and the experimental chemical shift ($\Delta = \delta_{\text{pred}} - \delta_{\text{exp}}$).

## Quantifying protonation states

A major advantage of the QNC-NMR protocol over static PBE and PBE0 approaches is its ability to provide an accurate description of fractional protonation states, which are inherently dynamic and cannot be captured within a static electronic-structure framework. Achieving such accuracy critically depends on a faithful representation of the underlying interatomic potential, as subtle changes in hydrogen bonding and proton sharing directly modulate the local magnetic environment. We therefore evaluate the performance of QNC-NMR by comparison with experimental data, focusing on observables that are directly sensitive to proton dynamics. Specifically, we examine the temperature dependence of $^1$H chemical shifts, which serve as a direct probe of the evolving hydrogen environment, as well as $^{15}$N chemical shifts, which provide an indirect but sensitive measure of the detailed composition of the protonation state.

The temperature-dependent chemical shift of the OH proton in the phenol-pyridine adduct (3-Methyl-4-nitro-phenol / 4-di-methyl-amino-pyridine) is shown in Figure 3a.[51,52] The shift, previously measured by Štoček et al.,[52] varies monotonically by 1 ppm as the temperature is raised from 173 to 348 K. This was ascribed to a change in protonation state of the intramolecular O-H···N interaction between both components in the cocrystal as the molecule progressively moves from a salt form to a co-crystal as the temperature increases.

Figure 3a shows how the QNC-NMR protocol correctly reproduces the temperature dependence of the protonation state in the structure.

Moreover, $^{15}$N chemical shifts are also sensitive probes of protonation states,[53] which makes it possible to determine the protonation of an N-H acceptor site, and this has been applied widely.[54] This high sensitivity makes $^{15}$N chemical shift predictions a compelling validation of our framework.

The QNC-NMR protocol also correctly reproduces the $^{15}$N chemical shifts for most of the compounds in the benchmark dataset of Ramos et al.[50] (Figure S2). Here two failure cases, GEHHIL (shown in Fig. 3b) and GEHHEH, are worth noting in more detail. The predicted chemical shielding of the nitrogen atom in the Schiff base moieties of GEHHIL and GEHHEH differ by 150 ppm depending on the protonation state of the base motif, as the shared proton exchanges between the base and a neighboring phenolic hydroxyl group.

The relative free energy of the two protonation states, that can be inferred straightforwardly from the probability of observing them in simulations, determines the ensemble-averaged chemical shieldings. The built-in uncertainty estimator in PET-MOLS allows us to estimate the uncertainty in the computed free-energy profile as a function of the predicted chemical shift of the Schiff base nitrogen in GEHHIL (Fig. 3c). This reveals that the ML potential, despite its accuracy, cannot

reliably resolve the subtle balance between the two protonation states due to the uncertain estimate of the free energy difference between the protonated and deprotonated base from PET-MOLS. Note that in Fig. 3c we use the chemical shift as the collective variable along which we visualize the free-energy profile, consistent with the general definition of free-energy surfaces in terms of a collective variable that can be any function of the atomic coordinates. Here, choosing the chemical shift instead of a more traditional geometric collective variable (e.g., the proton–acceptor distance or a proton-transfer coordinate) allows us to directly associate the free-energy uncertainty with a specific chemical-shift range. This example demonstrates the ability of our framework to self-diagnose, and to provide a qualitative aid to the interpretation of experiments, even in cases that are beyond its current energy resolution.

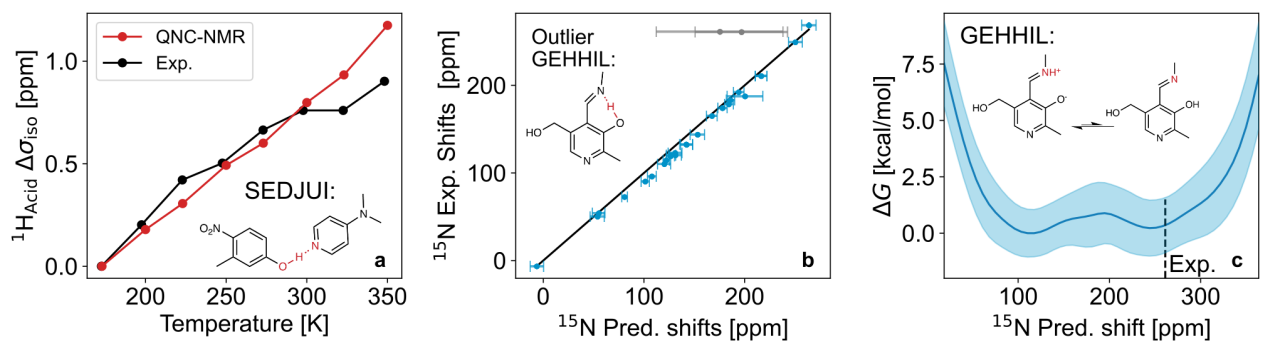

**Figure 3.** ShiftML3 predictions for the different protonation states. (a) Comparison of the $^1$H experimental (data reproduced from ref [52]) chemical shifts as a function of temperature with the QNC-NMR predictions in the 3-Methyl-4-nitro-phenol–4-di-methyl-amino-pyridine adduct (with the crystal structure deposited as entry SEDJUI[51] in the CCDC). (b) $^{15}$N Chemical shieldings compared to experimental benchmark values computed by ShiftML3 using the QNC-NMR approach. Error bars indicate the uncertainties in potential energy of the PET-MOLS propagated to chemical shieldings. We note the two outliers correspond to the nitrogens in the Schiff Bases of structures GEHHIL (shown inset) and GEHHEH (not shown) where QNC-NMR has high uncertainty of the protonation state (acidic proton highlighted in red). The diagonal black line marks the unit line. (c) Predicted free energy surface of the GEHHIL Schiff base protonation states as a function of the ShiftML3 computed $^{15}$N chemical shifts. PET-MOLS propagated uncertainties to the free energy are shown as bands corresponding to one standard deviation of the free energy.

## Amorphous structures

We note that studies of small crystalline systems, such as the ones presented in this work, would also be feasible using explicit *ab-initio* calculations, albeit extremely time consuming. They would be however entirely impossible when studying amorphous molecular solids that require models containing thousands of atoms to achieve an accurate description of the distribution of structural features.

As an illustration, we look at the GLP-1RA (**1-Cl**), shown in Figure 4a, constituted of 75 atoms and using structure models for the amorphous solid containing up to 320 molecules. Calculations of the $^1$H chemical shifts based on a combination of ShiftML3 and sampling of the amorphous structures using an empirical forcefield (FF-MD) show a large discrepancy for proton shifts of the acidic OH (H42) proton with experiment.[20] Specifically, Figure 4b shows the experimental distribution peaks at 13.2 ppm, while the predictions from the structures contained in the FF-MD simulations show a distribution centered at about 8 ppm. To resolve this discrepancy, we apply the PET-MOLS-based sampling protocol using 3 mutated versions of the compound (see SI for the mutation protocol used; mutations were used to substitute the two Cl atoms in the structure, since Cl is outside the current space of PET-MOLS).

Predictions for that labile site using the QNC-NMR protocol for the mutated structure (**1-H**), shown in Figure 4b, now display a marked bimodal separation between hydrogen-bonded and non–hydrogen-bonded environments. For labile protons in hydrogen bonds, the predicted shifts move downfield by about 5 ppm with a mean of 12.6 ppm relative to those obtained from the original FF-MD snapshots.

Figures 4c and 4d show ensemble-averaged predicted chemical shifts using the QNC-NMR protocol as a function of the average proton-transfer coordinate, $v = d(O-H_{42}) - d(H_{42} \cdots A)$ for all the local environments that are hydrogen-bonded for more than 50% of the time over the PIMD sampling using the PET-MOLS. The first observation in Figure 4c is a clear shift of the proton-transfer coordinate in the QNC-NMR approach towards higher values (i.e. stronger H-bonding) compared to the original FF-MD. This is due to the move from a force field based on harmonic potentials towards the incorporation of anharmonic behavior in the inter- and intramolecular interactions from the PET-MOLS trained on electronic calculations. This structural change translates into the dependence of the averaged predicted chemical shift along the trajectories as a function of $v$, as shown in Figure 4d, where we see a clear increase in the chemical shift. The restriction of the harmonic approximations using the FF based structures limits most of the predicted shifts to around 8 ppm, while the incorporation of finite-temperature and NQE shifts most of

the predictions to around 12.6 ppm, in agreement with Figure 4b.

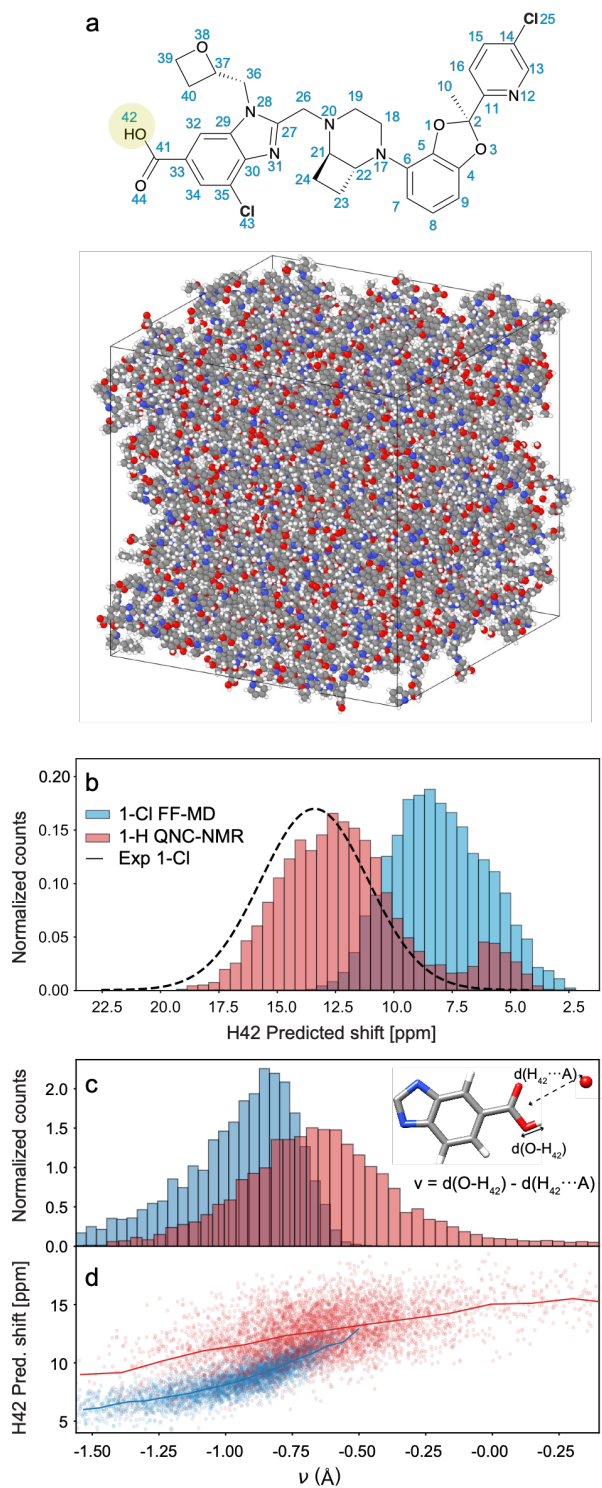

**Figure 4.** (a) Chemical structure of the GLP-1RA (**1-Cl**) and a snapshot of the amorphous atomistic model employed in this study. (b) Predicted chemical shift distribution for all the molecular environments for **1-Cl** from 17 snapshots selected from the FF-MD trajectories (blue), and the distribution of predicted shifts using the QNC-NMR protocol for each local molecular environment for the **1-H** mutant (red), compared to the experimentally determined chemical shift distribution (dashed black line) for the labile OH proton (H42 in (a)). The experimental distribution is taken from ref [20]. (c) Normalized distributions as a function of the proton-transfer coordinate, $v$, for all the hydrogen-bonded atomic OH environments obtained from FF-MD (for **1-Cl**) (blue) and from QNC-NMR (for **1-H**) (red). (d) Predicted $^1$H chemical shifts as a function of $v$ for the structures in (c). Solid lines represent the mean chemical shift computed within bins of $v$. In (c) and (d) hydrogen-bonded sites are defined as local atomic environments that participate in a hydrogen bond for more than 50% of the time over the PIMD sampling employing PET-MOLS.

This example provides a striking demonstration of how QNC-NMR crystallography, driven by fast ML surrogate models, provides a qualitatively superior description of H bonding relative to empirical forcefields, and how it can be applied to the size-scales necessary to investigate amorphous solids.

### Finetuning from experiment

The ability to sample a representative ensemble of structures consistent with the experimental conditions opens the possibility of refining the shielding model by finetuning it against experiment. To do so requires reconciliation of the ML model for shieldings, that makes predictions for a specific geometry $A$, and experiment, that measures averages over an ensemble of structures $\{A_i\}$. Using the fact that the last prediction step in our model involves a scalar product between learned features $\Phi(A)$ and a vector of weights $w$ (see SI for details), we can sample thousands of thermalized structures from PIMD simulations at experimental conditions, and from these compute an ensemble of latent features (ELF), $\{\Phi_i = \Phi(A_i)\}$. A corrected weight vector $w_{\text{corr}}$ can then be determined by a ridge regression in which the average ELF features are used to predict the experimental shieldings (see Methods). Reducing the complexity of the finetuning problem to a linear model, avoiding overfitting and backpropagating through the entire simulation trajectory when finetuning the ShiftML3 last layer weights, is crucial in light of the limited amount of well curated benchmark experimental data available to us (e.g. 19 compounds for a total of 123 $^{13}$C shielding values) and simulation trajectories containing tens of thousands images. We choose to demonstrate the finetuning via ELF on $^{13}$C shielding predictions rather than $^1$H because, for our purposes, the $^{13}$C benchmark set originally proposed by Ramos et al[50] constitutes to our knowledge the most diverse benchmark set of experimentally measured solid state chemical shifts with unambiguous assignments.

Within the ELF framework we first compute ensemble averages of the ShiftML3 last layer descriptors from PIMD simulations employing PET-MOLS, as well as computing ensemble averages of chemical shieldings using ShiftML3 for the set of 19 reference structures. By minimizing the differences between the computed ensemble and experimental shifts through a $\Delta$ model finetuning of the last layer weights of ShiftML3 we obtain an experimentally fine-tuned model of ShiftML3 for the prediction of $^{13}$C chemical shieldings. Even with this limited data, the improvement brought by an

experimentally-tuned Δ-ShiftML model is clear, with a reduction of about 0.1 ppm of the RMSE at the largest finetuning set size (Figure 5). Note that, as described in SI, we performed additional tests by either considering fixed or variable regression slopes in preparing the shifts for the Δ-model, as well as removing structure SUCROS04 whose GIPAW/ShiftML computed shifts deviate most from experimental values. Neither change the conclusions of this exercise, demonstrating that the improvement is not due to a few outliers but rather to an overall shift of the error distribution, as shown in the inset of Figure 5.

The Δ-model approach presented here is particularly helpful when learning from data obtained at inconsistent experimental conditions such as varying temperatures, or even mixed liquid and solid-state experiments. The ensemble-average formalism allows to incorporate and mix these conditions during the learning process by simply recreating the experimental conditions in the simulations of the ensemble-average descriptors. Extrapolating from the learning curve shown in Fig. 5, one can estimate that by using shifts from only a few hundred experimental structures would lead to a substantial improvement of the resolving power of NMR crystallography. Generating and making available this data requires a coordinated effort of the solid-state NMR community, and the authors hope that these results may encourage such an endeavor.

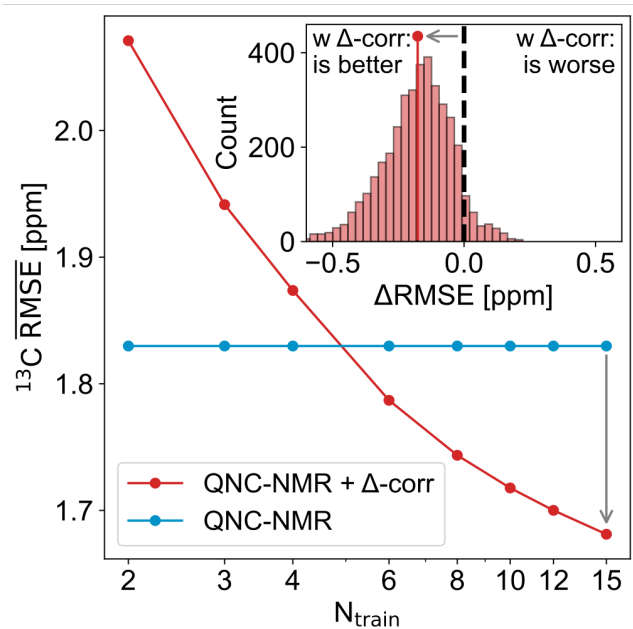

**Figure 5.** Learning curve of the last layer Δ-correction model for $^{13}$C chemical shieldings. The blue curve shows the QNC-NMR $^{13}$C reference prediction accuracy, the red curve shows the accuracy of experimentally corrected QNC-NMR predictions. The inset shows a histogram of the improvement in prediction accuracies, ΔRMSE, between uncorrected and corrected predictions considering all $\binom{19}{4}$ splits for the given 19 benchmark structures and fixed test set size of 4 structures. To the left of the dashed line in the inset are training splits for which the test set predictions of the experimentally corrected QNC-NMR shieldings are more accurate, and to the right are those for which the corrected predictions are less accurate.

## DISCUSSION

We have shown how the combination of the generally applicable ML potential PET-MOLS and chemical shielding models enable computational predictions of NMR experiments with unprecedented accuracy and scale. The more accurate electronic-structure reference, as well as the incorporation of finite-temperature and quantum delocalization effects capture the behavior of H-bonded protons without the need for empirical corrections. The result is a factor two reduction (from 1.63 to 0.75 ppm) in the RMSE between $^{1}$H shifts predictions and experiment for H-bonded protons. The availability of inexpensive uncertainty quantification increases further the reliability of the approach revealing the cases in which predictions cannot be trusted. The favorable scaling with system size allows us to show similar improvements for amorphous materials, that are otherwise inaccessible to explicit DFT simulations and constitute one of the most exciting areas of application of NMR crystallography.

The possibility of sampling accurately and at scale the quantum thermodynamic ensemble of molecular materials opens up the opportunity of crossing the boundary between theory and experiments. We demonstrate an ensemble of latent features scheme that can fine-tune a chemical shielding model against the very limited available experimental data, improving the accuracy of NMR structural determination beyond the limits of electronic structure calculations.

## METHODS

### Machine learning interatomic potential

We construct a machine learning potential for organic crystals at the PBE0+MBD DFT level of theory based on the point-edge transformer architecture (PET) from Pozdnyakov and Ceriotti.[55] This ML potential is referred to as PET-MOLS in this work. The training data was selected via farthest point sampling (FPS)[56] from the Cambridge Structural Database (CSD)[48] to cover a broad range of structural motifs, with the aim of making the potential broadly applicable and transferrable across molecular solids. The pool of candidates of the CSD database was limited to structures containing at most the elements H, C, N, O and up to 200 atoms to restrict memory and computational costs of reference computation. Experimental crystal structures were geometry relaxed using PBE-D2 and for each unique composition four thermally distorted structures were generated from short molecular dynamics simulations at 300 K using PBE-D2. The final training dataset consisted of 2,013 unique organic crystal compositions with a total of 10,056 training structures computed at the PBE0 level of

theory[57] with many-body dispersion (MBD).[58] Further information about reference DFT calculation, and the machine learning training of the model are reported in Section S3 in the SI. All computational experiments were performed without further finetuning of the PET-MOLS, underlining the transferability of the proposed MLIP across the space of organic solids.

### Reference computations
Reference PBE0+MBD calculations were performed with FHI-AIMS (version 210716.2) using "intermediate" 2020 default basis set definitions and a k-point density of 0.05 Å$^{-1}$. Dispersion corrections were computed using non self-consistent MBD@rsSCS corrections in their original formulation. Reference FHI-AIMS input and output files are made available in the Materials Cloud archive, associated with this publication.

### Experimental benchmark of crystalline structures and NMR predicted chemical shifts
We first evaluate the accuracy of $^1$H chemical shift predictions from static and QNC-NMR computations against a set of experimental benchmark structures from previous ShiftML publications.[34,35] The same evaluation was done for $^{13}$C and $^{15}$N experimental benchmarks from Ramos et al.[50] Only structures containing H, C, O, and N were included, consistent with the elemental coverage of the PET-MOLS training set.

In this study, we use the ShiftML3 machine learning model to predict chemical shieldings.[37] Shieldings σ are commonly converted to chemical shifts δ using a linear relation, $\delta = a\sigma + b$; and the prediction accuracy is evaluated via the root-mean-square error (RMSE) with respect to the experimental benchmark. In the best case, the fitting equation yields a slope of a = -1 indicating, that there is no systematic error, and this is the approach considered in this work for both crystalline and amorphous solids. Nonetheless, in literature fitting the slope has been shown to partially cancel out systematic errors, hence further discussion about fitted slope on the crystalline experimental benchmarks is discussed in Section S1 in the SI.

### Amorphous structures
The amorphous structure of the glucagon-like peptide-1 receptor agonist (GLP-1RA, **1-Cl**) recently published in our group[20] marks an interesting case for our approach as the labile proton site exhibit large deviations between predicted and experimental shifts distributions. This is not an isolated case as a systematic offset has also been observed on those local environments[18] responsible of the amorphous stabilization.[17-19]

We focus on correcting labile proton chemical shift predictions via the QNC-NMR protocol in amorphous structures of a series of chemically mutated derivatives of a GLP-1RA (**1-Cl**) agonist. The reasons of that choice are that the training set for PET-MOLS does not contain Cl atoms, and that only about 1% of those sites were acting as acceptors in the hydrogen-bond network responsible of the amorphous stabilization. Specifically, we consider analogues in which the chlorine substituent is replaced by hydrogen (**1-H**), a methyl group (**1-Me**), or a nitrile (**1-CN**). In Section S2 we study the acceptors counts for the different mutants showing negligible effects of the mutations on the hydrogen bond network after geometry optimization using PET-MOLS.

In this work, 17 MD random snapshots among all the reference MD generated ones in the original work[20] are selected as a test case.

### Reference molecular dynamics – amorphous structures
**1-Cl** and solvent molecules were optimized at the B3LYP-D3/6-31G** level, and CHELPG partial charges were derived from the corresponding electrostatic potentials. The 1-Cl, n-heptane and 2-MeTHF were parametrized using OPLS-2005, while water was modeled with TIP3P. Three systems were constructed – pure amorphous, C1c and C2b – using the Amorphous Cell module in Materials Studio. The composition of the solvent in the last two systems mention was varied while keeping the composition of the system the closest to experimental. Nonetheless, all systems contain 320 **1-Cl** structures. Then, five amorphous cubic cells were generated per system using COMPASS-II to provide consistent starting configurations.

GROMACS 2016.4 was used to run molecular simulations on all five independent replicas per system. Each trajectory underwent energy minimization, followed by a 1 ns NVT equilibration at 298 K using a Berendsen thermostat τ= 0.1 ps). Production runs consisted of 1 μs NPT simulations at 298 K and 1 atm using the stochastic velocity-rescale thermostat (τ = 0.1 ps) and the Parrinello–Rahman barostat (τ = 4 ps). PME was applied for electrostatics with a 1 nm cutoff, also used for van der Waals interactions; long-range dispersion corrections were included. All bonds involving hydrogen were constrained with LINCS, enabling a 2 fs timestep. Periodic boundary conditions were applied in all directions, and configurations were saved every 10 ps.

Only the last 200 ns of each replica were used for postprocessing; equilibration was assessed by monitoring the potential energy and the density as a function of simulation time. One snapshot every 100 ps was extracted from molecular trajectories, corresponding to 2000 snapshots per replica and 10,000 snapshots in total for each system.

### Path integral molecular dynamics
Path-integral molecular dynamics (PIMD) simulations were performed with i-PI (version 3.1)[59] using 32 beads in the NVT ensemble at the respective experimental temperatures. For the organic crystal benchmark set, simulations were run for 50 ps with a timestep of 0.5 fs, and a local PILE-L thermostat[60] with a damping time of 100 fs used for thermostating the centroid and a coupling parameter λ = 0.1 for the other normal modes, using 32 beads in the NVT ensemble at the respective experimental temperatures.

For the amorphous systems, snapshots from classical force-field (FF-MD) trajectories were first relaxed using the PET-MOLS via the metatomic LAMMPS interface,[61] followed by a pre-equilibration for ~200 ps running a classical NVT MD at 300 K using PET-MOLS and finally a 2ps PIGLET-accelerated-PIMD simulation using 8-beads.[62] In Section S2 it is shown the effect on structural changes in the hydrogen bond network for the Static-PBE0 geometries and the evolution of the averaged acidic H42 predicted shift along the simulations. All PIMD simulations of the experimental benchmark set were run at the respective experimental temperature. For the averaged chemical shift only the last ps of the PIMD was considered.

### Uncertainty quantification of thermodynamic averages

The PET-MOLS is equipped with a last layer ensemble uncertainty estimator, predicting efficiently a committee of energies for each structure.[63,64] The spread, or standard deviation of this committee of energy prediction serves as an estimate for the uncertainty of the predicted potential energy. The committee uncertainty estimates can then be propagated to a derived quantity, such as PIMD averages of chemical shifts, via thermodynamic reweighting, which avoids having to run multiple molecular dynamics simulations. Instead, a postprocessing step generates a committee of reweighted thermodynamic averages from a single MD trajectory. In practice a statistically more stable, cumulants based expansion is chosen for reweighting:[65]

$$\langle y \rangle_{V^{(k)}} \approx \langle y \rangle_{\bar{V}} - \beta \left[ \langle y(V^{(k)} - \bar{V}) \rangle_{\bar{V}} - \langle y \rangle_{\bar{V}} \langle V^{(k)} - \bar{V} \rangle_{\bar{V}} \right] \quad (1)$$

Here, $\langle y \rangle_{\bar{V}}$ is the canonical ensemble average of an observable y taken under the committee mean potential (sampling using PET-MOLS), $\beta$ is the thermodynamic beta, $V^{(k)}$ is the potential energy prediction of the committee member $k$, $\bar{V}$ is the mean potential energy prediction of the committee and $\langle y \rangle_{V^{(k)}}$ is approximately the canonical ensemble average for a given configuration, obtained after applying (1) instead of explicit molecular dynamics sampling using $V^{(k)}$.

The free energy surface (FES) is the logarithm of a partially integrated partition function, obtained by marginalizing over all microscopic degrees of freedom q. In practices, this reduces the number of remaining degrees of freedom according to arbitrarily defined collective variables with lower dimensionality, that maps a configuration q to a selected number of variables $z = \chi(q)$. This variable can also be the ShiftML3 predicted chemical shielding, and the resulting chemical shielding FES can therefore be expressed as the negative logarithm of $p(\sigma_{iso})$ the probability of finding a configuration at a given shielding value:

$$F(\sigma_{iso}) = -k_B T \ln(p(\sigma_{iso})) \quad (2)$$

In practice $p(\sigma_{iso})$ can be computed from MD and PIMD simulations via histograms or kernel-density estimates (KDE).

### Learning from experiments

We adopt a last-layer finetuning, writing ShiftML3 predictions as $\omega^\top \Phi(A_i)$ a product between a linear set of weights $\omega$ and a feature map $\phi(A_i)$ that transforms the local atomic environment $A_i$ into a fixed-size descriptor. Adapting this linearization, ensemble-averaged ShiftML predictions are straightforward to compute, and determining the corrected weights $\omega_{corr}$ that minimizes expression (3) reduces to a linear regression problem:

$$\Delta \sigma = \left( \sigma_{iso,exp.} - \omega_{corr}^\top \langle \Phi(A_i) \rangle \right)^2 \quad (3)$$

Due to the rather small number of training samples, and its conceptual simplicity, we minimize (3) using a delta model that corrects QNC-NMR predictions. The final corrected predictions are then obtained as $\sigma_{corr} = \sigma_{QNC\text{-}NMR} + \omega_{corr}^\top \langle \Phi(A_i)_{sML3} \rangle$, where $\langle \Phi(A_i)_{sML3} \rangle$ denotes the canonical ensemble average of the ShiftML3 last layer descriptor. We employ a ridge regressor and fixed regularization strength $\lambda=10$ to obtain $\omega_{corr}$. We split the 19 benchmark crystals into a training set of an increasing number of structures and keep always 4 test structures. Last layer ensemble averaged descriptors as well as averaged ShiftML3 shieldings are obtained from 50 ps PIMD simulations of each crystal at 300K using 32 beads. We repeat the model fitting across 2,000 random splits of the 19 crystals and compare the ShiftML3 thermal average shielding accuracies on the 4 held out structures against the Δ-model corrected values.

### DATA AVAILABILITY

We make the curated PET-MOLS training data, and example scripts to rerun the calculations of this work available via a materials-cloud archive associated with this publication made available upon publication.

### SUPPORTING INFORMATION

The supporting information is available free of charge upon publication.

### CORRESPONDING AUTHORS

lyndon.emsley@epfl.ch
michele.ceriotti@epfl.ch



\maketitle
($\square$) These authors contributed equally.\\
Raw data: All data are available from [xxx]

\tableofcontents
\clearpage

\setcounter{table}{0}
\setcounter{equation}{0}

\setcounter{figure}{0}
\setcounter{table}{0}
\renewcommand{\thefigure}{S\arabic{figure}}
\renewcommand{\thetable}{S\arabic{table}}

\setcounter{section}{0}
\renewcommand{\thesection}{S\arabic{section}}
\renewcommand{\thesubsection}{S\arabic{section}.\arabic{subsection}}
\renewcommand{\thefigure}{S\arabic{figure}}
\renewcommand{\thetable}{S\arabic{table}}

\IfFileExists{si-body.tex}{\input{si-body}}{%
  \IfFileExists{si-body}{\input{si-body}}{%
    \section*{Supplementary Information}
  }%
}

%

\clearpage



\printbibliography[title={Supplementary References}]

\end{document}

%% file: preamble.tex

\newcommand{\petpotential}{PET-MOLS}

%% file: si-body.tex
\section{Experimental benchmarks for the crystalline solids} \label{sec:experimentla_bench}

The structures for experimental benchmarks were compiled from Cordova et al \cite{cordova_machine_2022} for the $^1$H dataset while the $^{13}$C and $^{15}$N datasets of structures where compiled from Ramos \textit{et al.} \cite{ramosInterplayDensityFunctional2024}. These structures were used without further relaxation for chemical shift prediction with ShiftML3 \cite{kellnerDeepLearningModel2025} for the Static-PBE benchmark. As reported, $^1$H experimental benchmark are all atoms geometries relaxed using D2 dispersion corrections keeping the cell parameters fixed.  The $^{13}$C and $^{15}$N used the dispersion correction D3(BJ) for geometry optimization of all atoms keeping the cell parameters fixed.

\rev{\subsection{Experimental shifts and predicted shieldings for the \texorpdfstring{$^{1}$H}{1H} dataset}}

Assigned chemical shift values were used from the previously curated data set. \cite{cordova_machine_2022} The methyl rotational dynamics were accounted for by averaging the chemical shifts of the three proton positions into a single effective value per methyl group, which was treated as contributing threefold shielding to reflect the three equivalent protons. The same approached is done for magnetically equivalent protons (i.e. methylene groups). In cases of pairwise ambiguities, experimental chemical shifts were resolved by selecting the assignment yielding the lowest shift RMSE. (i.e. H$_a$ and H$_b$ of CH$_2$ groups)

{
\renewcommand{\arraystretch}{0.5}

\begin{longtable}{ccccc}
\caption{Experimental chemical shifts and ShiftML3 predicted shieldings for the \rev{209} atomic sites in the $^{1}$H benchmark dataset of structures from ShiftML2 \cite{cordova_machine_2022}\rev{.} Predicted shieldings are given for the Static-PBE,  Static-PBE0 geometries, and QNC-NMR ensemble.}\label{tab:1H_tables} \\
\toprule
\multirow{2}{*}{\textbf{Ref. Code}} &
\multirow{2}{*}{\textbf{Experimental shift [ppm]}} &
\multicolumn{3}{c}{\textbf{Predicted shieldings [ppm]}} \\
\cmidrule{3-5}
& & \textbf{Static-PBE} & \textbf{Static-PBE0} & \textbf{QNC-NMR} \\
\midrule
\endhead

    \textbf{Atuliflapon} & 1.20   & 29.32 & 29.49  & 28.72 \\
          & 1.20   & 29.32 & 29.49  & 28.72 \\
          & 1.20   & 29.32 & 29.49  & 28.72 \\
          & 10.60  & 18.11 & 18.15 & 18.09 \\
          & 5.80   & 23.78 & 24.03 & 23.49 \\
          & 6.90   & 23.88 & 23.99 & 23.61 \\
          & 7.30   & 23.06 & 23.23 & 22.92 \\
          & 6.70   & 24.34 & 24.57 & 24.14 \\
          & 7.00    & 22.78 & 22.94 & 22.54 \\
          & 3.90   & 27.01 & 27.26 & 26.53 \\
          & 1.60   & 28.67 & 28.90 & 27.98 \\
          & 0.00   & 30.35 & 30.52 & 29.62 \\
          & 1.70   & 28.55 & 28.75 & 27.98 \\
          & 1.60   & 28.90 & 29.37 & 28.45 \\
          & 1.60   & 29.01 & 29.06 & 28.34 \\
          & -0.50  & 30.02 & 30.19 & 29.28 \\
          & 0.80   & 29.51 & 29.72 & 28.84 \\
          & -0.50  & 30.57 & 30.72 & 29.82 \\
          & 0.80   & 29.26 & 29.41 & 28.72 \\
          & 7.70   & 21.36 & 21.72 & 21.25 \\
          & 7.60   & 22.58 & 22.88 & 22.41 \\
          & 1.70   & 29.63 & 29.61 & 28.52 \\
          & 2.70   & 27.80  & 27.94 & 27.26 \\
          & 6.90   & 22.46 & 23.41 & 22.81 \\
          & 1.90   & 28.07 & 28.24 & 27.43 \\
          & 2.70   & 27.50  & 27.64 & 26.88 \\
    \midrule
    \textbf{AZD8329 Form 4} & 6.92  & 24.13 & 24.35 & 23.97 \\
          & 8.69  & 22.30  & 22.54 & 22.30 \\
          & 9.01  & 21.50  & 21.49 & 21.39 \\
          & 8.47  & 22.80  & 22.84  & 22.70 \\
          & 15.37 & 13.65 & 15.90 & 15.30 \\
          & 7.73  & 22.43 & 22.59 & 22.12 \\
          & 9.64  & 19.60  & 20.65 & 20.21 \\
          & 2.90   & 28.11 & 28.23 & 27.39 \\
          & 1.78  & 28.80  & 28.95 & 28.07 \\
          & 1.88  & 29.47 & 29.69 & 28.69 \\
          & 1.88  & 28.47 & 28.60 & 27.92 \\
          & 1.80   & 28.86 & 29.03 & 28.18 \\
          & 1.60   & 29.58 & 29.86 & 29.03 \\
          & 0.44  & 30.29 & 30.48  & 29.54 \\
          & 1.54  & 29.01 & 29.08 & 28.28 \\
          & 1.88  & 28.62 & 29.02 & 28.24 \\
          & 1.88  & 29.54 & 29.68 & 28.83 \\
          & 0.80   & 29.83 & 29.97 & 29.11 \\
          & 0.80   & 30.11 & 30.18 & 29.27 \\
          & 1.00    & 29.17 & 29.30 & 28.44 \\
          & 1.74  & 29.03 & 29.29 & 28.48 \\
          & 1.74  & 29.75 & 29.92 & 29.13 \\
          & 0.73  & 30.09 & 30.37 & 29.57 \\
          & 0.73  & 30.09 & 30.37 & 29.57 \\
          & 0.73  & 30.09 & 30.37 & 29.57 \\
          & 0.73  & 29.90  & 30.08  & 29.20 \\
          & 0.73  & 29.90  & 30.08  & 29.20 \\
          & 0.73  & 29.90  & 30.08  & 29.20 \\
          & 0.73  & 30.76 & 31.12 & 29.98 \\
          & 0.73  & 30.76 & 31.12 & 29.98 \\
          & 0.73  & 30.76 & 31.12 & 29.98 \\
    \midrule
    \textbf{Cocaine} & 3.76  & 27.01 & 27.17 & 26.32 \\
          & 3.78  & 27.24 & 27.47 & 26.89 \\
          & 5.63  & 25.44 & 25.63 & 24.87 \\
          & 3.32  & 27.33 & 27.53 & 26.80 \\
          & 3.06  & 28.32 & 28.44 & 27.81 \\
          & 3.49  & 28.06 & 28.36 & 27.63 \\
          & 2.91  & 28.98 & 29.19 & 28.28 \\
          & 3.38  & 28.33 & 28.60 & 27.79 \\
          & 2.56  & 28.64 & 28.83 & 28.06 \\
          & 2.12  & 28.97 & 29.13 & 28.40 \\
          & 1.04  & 29.47 & 29.62 & 28.70 \\
          & 1.04  & 29.47 & 29.62 & 28.70 \\
          & 1.04  & 29.47 & 29.62 & 28.70 \\
          & 8.01  & 22.53 & 22.75 & 22.35 \\
          & 8.01  & 22.99 & 23.13 & 23.02 \\
          & 8.01  & 22.79 & 22.94 & 22.60 \\
          & 8.01  & 22.73 & 22.91 & 22.52 \\
          & 8.01  & 22.86 & 23.04 & 22.68 \\
          & 3.78  & 26.81 & 27.03 & 26.31 \\
          & 3.78  & 26.81 & 27.03 & 26.31 \\
          & 3.78  & 26.81 & 27.03 & 26.31 \\
    \midrule
    \textbf{Dimethylimidazole} & 4.80   & 25.91 & 25.92 & 25.09 \\
          & 0.70   & 29.53 & 29.87 & 29.04 \\
          & 0.70   & 29.53 & 29.87 & 29.04 \\
          & 0.70   & 29.53 & 29.87 & 29.04 \\
          & 1.40   & 29.41 & 29.57 & 28.77 \\
          & 1.40   & 29.41 & 29.57 & 28.77 \\
          & 1.40   & 29.41 & 29.57 & 28.77 \\
          & 13.00    & 16.30  & 16.10 & 16.46 \\
          & 1.40   & 29.08 & 29.31 & 28.57 \\
          & 1.40   & 29.08 & 29.31 & 28.57 \\
          & 1.40   & 29.08 & 29.31 & 28.57 \\
          & 1.50   & 28.97 & 29.15 & 28.32 \\
          & 1.50   & 28.97 & 29.15 & 28.32 \\
          & 1.50   & 28.97 & 29.15 & 28.32 \\
          & 15.00    & 14.23 & 14.54 & 14.47 \\
          & 5.20   & 24.83 & 24.96 & 24.39 \\
    \midrule
    \textbf{Flufenamic acid} & 8.30   & 22.36 & 22.36 & 22.10 \\
          & 6.00     & 24.58 & 24.79 & 24.28 \\
          & 5.40   & 25.46 & 25.65 & 25.07 \\
          & 6.80   & 23.33 & 23.63 & 23.27 \\
          & 12.40  & 16.04 & 16.94 & 16.21 \\
          & 9.60   & 19.59 & 19.97 & 19.43 \\
          & 6.90   & 23.59 & 23.66 & 23.23 \\
          & 6.20   & 24.08 & 24.25 & 23.79 \\
          & 5.90   & 24.62 & 24.93 & 24.35 \\
          & 7.30   & 23.85 & 24.00 & 23.52 \\
    \midrule
    \textbf{Flutamide} & 7.10   & 23.18 & 23.42 & 22.98 \\
          & 9.90   & 20.66 & 20.63 & 20.56 \\
          & 8.00     & 23.08 & 23.33 & 22.96 \\
          & 8.00     & 21.03 & 21.82 & 21.42 \\
          & 2.00     & 28.39 & 28.53 & 27.72 \\
          & 1.20   & 29.51 & 29.71 & 28.85 \\
          & 1.20   & 29.51 & 29.71 & 28.85 \\
          & 1.20   & 29.51 & 29.71 & 28.85 \\
          & 1.20   & 29.51 & 29.71 & 28.85 \\
          & 1.20   & 29.51 & 29.71 & 28.85 \\
          & 1.20   & 29.51 & 29.71 & 28.85 \\
    \midrule
    \textbf{Furosemide} & 8.40   & 22.47 & 22.76 & 22.29 \\
          & 12.70  & 16.73 & 17.10 & 15.64 \\
          & 6.50   & 23.80 & 24.05 & 23.64 \\
          & 6.50   & 23.80 & 24.05 & 23.64 \\
          & 7.90   & 23.26 & 23.26 & 22.86 \\
          & 8.70   & 22.41 & 22.54 & 22.19 \\
          & 5.60   & 25.59 & 25.54 & 24.97 \\
          & 3.80   & 26.57 & 26.60 & 25.83 \\
          & 6.00     & 24.26 & 24.55 & 24.08 \\
          & 6.00     & 24.40 & 24.64 & 24.15 \\
          & 7.70   & 22.47 & 22.64 & 22.33 \\
          & 8.40   & 22.39 & 22.59 & 21.99 \\
          & 4.30   & 25.70 & 25.81 & 25.19 \\
          & 12.70  & 16.43 & 17.13 & 15.73 \\
          & 6.70   & 22.98 & 23.45 & 23.15 \\
          & 6.70   & 22.98 & 23.45 & 23.15 \\
          & 6.20   & 25.23 & 25.47 & 24.86 \\
          & 8.60   & 22.45 & 22.58 & 22.30 \\
          & 4.30   & 25.64 & 25.88 & 25.24 \\
          & 5.70   & 24.64 & 24.91 & 24.38 \\
          & 6.40   & 23.48 & 23.67 & 23.33 \\
          & 6.50   & 23.76 & 24.10 & 23.66 \\
    \midrule
    \textbf{\makecell{Indomethacin-\\nicotinamide}} & 7.30   & 23.20 & 23.39 & 23.08 \\
          & 5.50   & 25.14 & 25.31 & 24.86 \\
          & 6.80   & 23.95 & 24.13 & 23.61 \\
          & 0.90   & 30.04 & 30.15 & 29.39 \\
          & 0.90   & 30.04 & 30.15 & 29.39 \\
          & 0.90   & 30.04 & 30.15 & 29.39 \\
          & 2.90   & 28.02 & 28.19 & 27.20 \\
          & 2.90   & 28.02 & 28.19 & 27.20 \\
          & 2.90   & 28.02 & 28.19 & 27.20 \\
          & 3.40   & 27.69 & 27.83 & 27.15 \\
          & 3.40   & 27.69 & 27.83 & 27.15 \\
          & 16.30  & 13.47 & 14.43 & 12.77 \\
          & 6.40   & 25.07 & 25.28 & 24.76 \\
          & 6.00     & 23.46 & 23.92 & 23.65 \\
          & 6.00     & 25.09 & 25.24 & 24.59 \\
          & 6.40   & 24.61 & 24.78 & 24.41 \\
          & 9.80   & 21.71 & 21.95 & 21.60 \\
          & 9.80   & 22.29 & 22.54 & 22.10 \\
          & 8.30   & 23.27 & 23.59 & 23.17 \\
          & 7.70   & 23.35 & 23.61 & 23.25 \\
          & 7.30   & 22.36 & 22.62 & 22.58 \\
          & 9.00     & 21.13 & 21.58 & 21.54 \\
    \midrule
    \textbf{Naproxen} & 7.00     & 23.09 & 23.29 & 22.95 \\
          & 6.10   & 24.24 & 24.32 & 23.91 \\
          & 3.80   & 26.96 & 27.07 & 26.31 \\
          & 4.50   & 25.69 & 25.97 & 25.41 \\
          & 4.10   & 25.85 & 25.95 & 25.43 \\
          & 5.90   & 24.93 & 25.05 & 24.58 \\
          & 3.20   & 27.73 & 27.79  & 26.96 \\
          & 1.80   & 29.36 & 29.57 & 28.75 \\
          & 1.80   & 29.36 & 29.57 & 28.75 \\
          & 1.80   & 29.36 & 29.57 & 28.75 \\
          & 2.30   & 28.25 & 28.44 & 27.49 \\
          & 2.30   & 28.25 & 28.44 & 27.49 \\
          & 2.30   & 28.25 & 28.44 & 27.49 \\
          & 11.50  & 16.32 & 17.36 & 17.06 \\
    \midrule
    \textbf{Penicillin} & 4.10   & 26.72 & 26.78 & 26.16 \\
          & 6.40   & 24.53 & 24.90 & 24.19 \\
          & 5.70   & 25.45 & 25.38 & 24.88 \\
          & 1.70   & 29.49 & 29.68 & 28.99 \\
          & 1.70   & 29.49 & 29.68 & 28.99 \\
          & 1.70   & 29.49 & 29.68 & 28.99 \\
          & 0.90   & 30.39 & 30.49 & 29.55 \\
          & 0.90   & 30.39 & 30.49 & 29.55 \\
          & 0.90   & 30.39 & 30.49 & 29.55 \\
          & 6.20   & 24.91 & 25.64 & 24.92 \\
          & 3.90   & 27.22 & 26.80 & 25.53 \\
          & 4.70   & 26.32 & 25.96 & 25.28 \\
          & 7.10   & 23.57 & 23.71 & 23.25 \\
          & 7.10   & 23.57 & 23.71 & 23.25 \\
          & 7.10   & 23.57 & 23.71 & 23.25 \\
          & 7.10   & 23.57 & 23.71 & 23.25 \\
          & 7.10   & 23.57 & 23.71 & 23.25 \\
    \midrule
    \textbf{\makecell{Phenylphosphonic\\acid}} & 12.70  & 16.84 & 17.75 & 16.76 \\
          & 11.50  & 18.20  & 19.38 & 18.48 \\
          & 6.50   & 24.74 & 24.94 & 24.39 \\
          & 6.80   & 24.16 & 24.38 & 23.99 \\
          & 8.50   & 22.28 & 22.54 & 22.27 \\
          & 7.10   & 24.11 & 24.32 & 23.77 \\
          & 6.20   & 24.82 & 24.98 & 24.44 \\
    \midrule
    \textbf{Theophylline} & 14.60  & 14.32 & 15.46  & 15.36 \\
          & 7.70   & 22.80  & 23.03 & 22.37 \\
          & 3.40   & 27.49 & 27.72 & 27.04 \\
          & 3.40   & 27.49 & 27.72 & 27.04 \\
          & 3.40   & 27.49 & 27.72 & 27.04 \\
          & 3.40   & 27.49 & 27.72 & 27.04 \\
          & 3.40   & 27.49 & 27.72 & 27.04 \\
          & 3.40   & 27.49 & 27.72 & 27.04 \\
    \midrule
    \textbf{Uracil} & 7.50   & 21.99 & 22.05 & 21.86 \\
          & 10.80  & 18.32 & 19.06  & 18.98 \\
          & 11.20  & 17.98 & 19.19 & 18.96 \\
          & 6.00     & 23.99 & 24.37 & 23.91 \\
    \bottomrule
\end{longtable}
}

\newpage
\rev{\subsection{Experimental shifts and predicted shieldings for the \texorpdfstring{$^{13}$C}{13C} dataset}}

{\renewcommand{\arraystretch}{0.5}
\begin{longtable}{lcccc}
\caption{Experimental chemical shifts and ShiftML3 predicted shieldings  for the 1\rev{32} atomic sites in the $^{13}$C benchmark dataset of structures from Ramos et al. \cite{ramosInterplayDensityFunctional2024}\rev{.} Predicted shieldings are given for the Static-PBE and Static-PBE0 geometries, and in the QNC-NMR ensemble. The shifts below are ordered according to their appearance in the CIF files.} \label{tab:table_13C} \\

\toprule
\multirow{2}{*}{\textbf{Ref. Code}} &
\multirow{2}{*}{\textbf{Experimental shift [ppm]}} &
\multicolumn{3}{c}{\textbf{Predicted shieldings [ppm]}} \\
\cmidrule{3-5}
& & \textbf{Static-PBE} & \textbf{Static-PBE0} & \textbf{QNC-NMR} \\
\midrule
\endhead
    \textbf{ADENOS12} & 154.10 & 13.01 & 17.45 & 14.64 \\
          & 147.90 & 19.54 & 22.86 & 20.41 \\
          & 119.30 & 46.12 & 49.73 & 46.35 \\
          & 154.70 & 15.15 & 18.71 & 16.76 \\
          & 137.40 & 28.15 & 32.45 & 28.68 \\
          & 91.90  & 75.88 & 80.10 & 75.88 \\
          & 74.60  & 92.11 & 96.35 & 91.87 \\
          & 70.90  & 96.01 & 99.77 & 94.92 \\
          & 84.50  & 81.59 & 86.75 & 81.87 \\
          & 62.50  & 106.96 & 110.85 & 105.08 \\
    \midrule
    \textbf{ASPARM03} & 176.40 & -12.11 & -4.34  & -6.27 \\
          & 51.80  & 119.42 & 122.90 & 117.44 \\
          & 36.10  & 136.95 & 138.63 & 132.46 \\
          & 177.10 & -7.22 & -1.44 & -3.98 \\
    \midrule
    \textbf{FRUCTO02} & 65.40  & 102.71 & 106.42 & 101.68 \\
          & 99.70  & 62.66 & 68.49 & 64.72 \\
          & 67.20  & 99.59 & 103.60 & 98.04 \\
          & 69.00    & 99.14 & 102.16 & 96.78 \\
          & 71.40  & 96.00  & 100.45 & 95.79 \\
          & 64.90  & 103.54 & 107.97 & 102.47 \\
    \midrule
    \textbf{GLUTAM01} & 173.00   & -9.86 & -2.40  & -4.88 \\
          & 53.30  & 115.99 & 119.47 & 114.21 \\
          & 25.50  & 145.36 & 147.31 & 140.89 \\
          & 28.50  & 143.23 & 145.42 & 138.79 \\
          & 176.50 & -8.84 & -3.10  & -6.08 \\
    \midrule
    \textbf{GLYCIN29} & 176.20 & -10.94 & -3.91  & -5.96 \\
          & 43.50  & 128.90 & 131.08 & 125.66 \\
    \midrule
    \textbf{HXACAN09} & 133.10 & 37.49 & 40.95 & 37.04 \\
          & 123.40 & 46.80  & 48.02 & 43.77 \\
          & 115.70 & 52.53 & 54.38 & 49.79 \\
          & 152.30 & 14.47 & 17.48 & 14.09 \\
          & 116.40 & 51.99 & 54.46 & 49.48 \\
          & 120.60 & 49.32 & 50.26 & 45.63 \\
          & 169.80 & -4.05 & 2.63  & 0.42 \\
          & 23.80  & 148.73 & 150.68 & 145.08 \\
    \midrule
    \textbf{LALNIN12} & 176.80 & -13.47 & -6.38 & -8.33 \\
          & 50.90  & 119.84 & 123.41 & 118.06 \\
          & 19.80  & 154.74 & 156.07 & 149.71 \\
    \midrule
    \textbf{LCYSTN21} & 174.00 & -9.71 & -1.74 & -4.64 \\
          & 56.70  & 114.10 & 117.20 & 112.25 \\
          & 28.80  & 139.78 & 144.15 & 138.38 \\
    \midrule
    \textbf{LSERIN01} & 175.10 & -10.53 & -3.11  & -5.20 \\
          & 55.60  & 113.87 & 116.86 & 111.95 \\
          & 62.90  & 104.28 & 108.32 & 103.51 \\
    \midrule
    \textbf{LSERMH10} & 175.60 & -12.45 & -5.20 & -7.10 \\
          & 58.30  & 111.98 & 115.28 & 110.54 \\
          & 61.80  & 106.96 & 110.45 & 104.90 \\
    \midrule
    \textbf{LTHREO01} & 171.90 & -5.76 & 1.01  & -1.84 \\
          & 61.20  & 110.40 & 113.44 & 108.01 \\
          & 66.80  & 98.49 & 102.68 & 97.03 \\
          & 20.40  & 152.89 & 153.95 & 148.03 \\
    \midrule
    \textbf{LTYROS11} & 175.40 & -12.34 & -5.35 & -7.51 \\
          & 130.30 & 36.97 & 39.01 & 35.23 \\
          & 116.40 & 52.18 & 54.64 & 50.30 \\
          & 56.40  & 114.49 & 117.87 & 112.99 \\
          & 131.00   & 36.84 & 38.51 & 34.59 \\
          & 118.00   & 51.96 & 54.47 & 49.67 \\
          & 155.60 & 11.30  & 14.05 & 10.36 \\
          & 36.80  & 134.78 & 136.58 & 130.98 \\
          & 123.60 & 44.23 & 46.40 & 41.57 \\
    \midrule
    \textbf{MBDGAL02} & 105.70 & 58.04 & 62.97 & 58.52 \\
          & 71.20  & 97.59 & 101.15 & 95.68 \\
          & 72.10  & 95.11 & 98.84 & 93.72 \\
          & 69.30  & 96.52 & 100.54 & 96.01 \\
          & 75.60  & 89.06 & 92.57 & 88.12 \\
          & 62.80  & 106.24 & 109.42 & 103.83 \\
          & 57.60  & 111.18 & 116.16 & 110.04 \\
    \midrule
    \textbf{MEMANP11} & 99.60  & 64.94 & 69.83  & 65.62 \\
          & 71.30  & 95.11 & 98.63 & 94.30 \\
          & 71.70  & 96.10  & 99.99 & 95.14 \\
          & 64.80  & 102.86 & 106.82 & 100.88 \\
          & 71.90  & 96.03 & 100.24 & 94.88 \\
          & 58.90  & 108.85 & 112.28 & 106.58 \\
          & 54.90  & 113.93 & 118.07 & 112.69 \\
    \midrule
    \textbf{MGALPY01} & 100.40 & 63.56 & 69.12  & 64.50 \\
          & 67.60  & 100.43 & 103.87 & 98.12 \\
          & 72.60  & 93.10  & 97.32 & 92.96 \\
          & 70.00  & 98.13 & 101.70 & 96.60 \\
          & 72.90  & 93.84 & 98.44 & 93.22 \\
          & 61.40  & 106.83 & 109.98 & 104.16 \\
          & 55.20  & 113.94 & 118.69 & 112.61 \\
    \midrule
    \textbf{MGLUCP11} & 101.00   & 63.38 & 68.85  & 64.40 \\
          & 72.30  & 95.72 & 99.02  & 93.74 \\
          & 74.60  & 93.19 & 97.36 & 92.63 \\
          & 72.50  & 93.47 & 97.20 & 92.63 \\
          & 75.30  & 94.69 & 97.94 & 92.29 \\
          & 63.80  & 105.34 & 108.94 & 103.95 \\
          & 56.50  & 114.09 & 118.56 & 112.68 \\
    \midrule
    \textbf{PERYTO10} & 50.20  & 120.38 & 121.48 & 116.49 \\
          & 58.40  & 109.49 & 113.24 & 107.63 \\
    \midrule
    \textbf{RHAMAH12} & 94.50  & 71.12 & 76.54 & 71.70 \\
          & 72.20  & 95.24 & 99.59 & 94.77 \\
          & 71.00  & 98.13 & 101.46 & 96.17 \\
          & 72.50  & 95.38 & 99.44 & 94.11 \\
          & 69.80  & 98.22 & 102.69 & 97.17 \\
          & 17.80  & 157.07 & 158.32 & 152.71 \\
    \midrule
    \textbf{SUCROS04} & 93.30  & 71.48 & 76.75 & 72.03 \\
          & 66.00    & 105.35 & 107.71 & 102.41 \\
          & 73.70  & 93.45 & 96.69  & 91.60 \\
          & 102.40 & 52.83 & 59.47 & 56.08 \\
          & 72.80  & 93.67 & 98.29 & 94.04 \\
          & 82.90  & 87.08 & 90.03 & 85.55 \\
          & 67.90  & 99.91 & 103.09 & 97.49 \\
          & 71.80  & 96.26 & 99.89 & 94.60 \\
          & 73.60  & 93.69 & 97.28 & 92.37 \\
          & 81.80  & 83.50  & 86.94 & 82.60 \\
          & 60.00  & 108.80 & 112.42 & 106.82 \\
          & 61.00  & 106.43 & 110.23 & 104.53 \\
    \midrule
    \textbf{SULAMD06} & 127.10 & 39.82 & 45.98 & 40.35 \\
          & 129.50 & 39.63 & 41.09 & 37.91 \\
          & 117.10 & 54.60 & 56.92 & 51.66 \\
          & 153.40 & 19.71 & 21.62 & 18.58 \\
          & 112.30 & 57.54 & 60.71 & 55.10 \\
          & 129.50 & 40.99 & 42.61 & 39.00 \\
    \midrule
    \textbf{TRIPHE11} & 126.40 & 42.10  & 44.26 & 39.54 \\
          & 129.50 & 39.33 & 42.49  & 39.17 \\
          & 124.50 & 44.94 & 48.04 & 42.93 \\
          & 125.90 & 41.24 & 43.52 & 39.55 \\
          & 127.50 & 42.78 & 44.38 & 40.01 \\
          & 122.30 & 48.91 & 50.63 & 45.53 \\
          & 130.20 & 38.92 & 40.77  & 37.81 \\
          & 129.50 & 40.70  & 41.57 & 38.78 \\
          & 120.90 & 50.31 & 51.74 & 45.99 \\
          & 125.90 & 44.70  & 45.03 & 40.81 \\
          & 121.70 & 49.19 & 50.25 & 45.44 \\
          & 129.50 & 40.67 & 42.28 & 39.28 \\
          & 129.50 & 39.67 & 40.87 & 38.12 \\
          & 122.30 & 48.50  & 50.08 & 45.30 \\
          & 126.90 & 41.66 & 43.79 & 39.52 \\
          & 126.90 & 42.10  & 43.06 & 38.80 \\
          & 123.80 & 44.47 & 48.65 & 44.02 \\
          & 129.80 & 40.07 & 41.05 & 38.36 \\
    \bottomrule

  \label{tab:13c-table}%

\end{longtable}
}
\newpage
\rev{\subsection{Experimental shifts and predicted shieldings for the \texorpdfstring{$^{15}$N}{15N} dataset}}

{\renewcommand{\arraystretch}{0.5}
\begin{longtable}{lcccc}
\caption{Experimental chemical shifts and ShiftML3 predicted shieldings for the \rev{35} atomic sites in the $^{15}$N benchmark dataset of structures from Ramos et al. \cite{ramosInterplayDensityFunctional2024} \rev{.} Predicted shieldings are given for the Static-PBE and Static-PBE0 geometries, and in the QNC-NMR ensemble.} \label{tab:table_15N} \\
\toprule
\multirow{2}{*}{\textbf{Ref. Code}} &
\multirow{2}{*}{\textbf{Experimental shift [ppm]}} &
\multicolumn{3}{c}{\textbf{Predicted shieldings [ppm]}} \\
\cmidrule{3-5}
& & \textbf{Static-PBE} & \textbf{Static-PBE0} & \textbf{QNC-NMR} \\
\midrule
\endfirsthead
\toprule
\multirow{2}{*}{\textbf{Ref. Code}} &
\multirow{2}{*}{\textbf{Experimental shift [ppm]}} &
\multicolumn{3}{c}{\textbf{Predicted shieldings [ppm]}} \\
\cmidrule{3-5}
& & \textbf{Static-PBE} & \textbf{Static-PBE0} & \textbf{QNC-NMR} \\
\midrule
\endhead
\textbf{BITZAF}   & 249.50 & -68.97 & -62.01 & -64.41 \\
\midrule
\textbf{GEHHEH}   & 187.40 &  11.16 &  13.25 & -13.80 \\
                  & 261.00 & -51.89 & -58.21 & -49.37 \\
\midrule
\textbf{GEHHIL}   & 268.50 & -81.30 & -74.39 & -78.54 \\
                  & 261.20 & -64.26 & -69.87 & -50.78 \\
\midrule
\textbf{LHISTD02} & 210.80 & -30.80 & -20.79 & -30.32 \\
                  & 132.60 &  45.81 &  52.48 &  43.69 \\
\midrule
\textbf{LHISTD13} & 210.60 & -30.58 & -20.50 & -30.35 \\
                  & 132.40 &  44.66 &  52.03 &  43.27 \\
\midrule
\textbf{TEJWAG}   & 143.90 &  38.03 &  42.86 &  31.13 \\
\midrule
\textbf{GLYCIN03} &  -6.50 & 195.64 & 199.57 & 191.39 \\
\midrule
\textbf{FUSVAQ01} & 183.20 &  -0.19 &   8.09 &   0.48 \\
                  & 174.20 &   8.87 &  15.23 &   8.59 \\
                  & 192.20 &  -9.81 &  -1.04 &  -7.99 \\
                  & 120.20 &  56.21 &  62.60 &  55.26 \\
                  &  50.20 & 131.18 & 136.42 & 131.70 \\
\midrule
\textbf{CYTSIN}   & 110.20 &  65.71 &  74.87 &  59.80 \\
                  & 165.20 &  17.78 &  25.14 &  22.23 \\
                  &  54.20 & 133.13 & 137.10 & 129.83 \\
\midrule
\textbf{THYMIN01} &  90.20 &  83.12 &  92.32 &  83.86 \\
                  & 119.20 &  58.68 &  65.91 &  59.26 \\
\midrule
\textbf{CIMETD}   & 130.50 &  45.40 &  52.01 &  44.47 \\
                  & 213.10 & -32.76 & -24.66 & -33.48 \\
                  &  56.60 & 125.63 & 133.37 & 124.05 \\
                  &  43.50 & 140.32 & 149.20 & 139.17 \\
                  &  45.80 & 139.41 & 146.13 & 138.84 \\
                  & 149.90 &  34.30 &  36.37 &  31.26 \\
\midrule
\textbf{BAPLOT01} & 114.70 &  59.89 &  68.39 &  63.18 \\
                  &  72.70 & 101.92 & 109.49 & 104.54 \\
                  & 122.70 &  53.71 &  63.58 &  55.32 \\
                  & 178.70 &   1.61 &   8.72 &   0.93 \\
\midrule
\textbf{LTYRHC10} &   8.00 & 166.90 & 184.36 & 174.66 \\
\midrule
\textbf{CYSCLM}   &   1.50 & 186.34 & 190.64 & 181.66 \\
\midrule
\textbf{URACIL}   &  96.20 &  75.63 &  84.10 &  76.83 \\
                  & 120.20 &  59.00 &  65.38 &  58.41 \\
    \bottomrule
  \label{tab:15Ntable}%
\end{longtable}
}
\subsection{Parity plots between experimental and predicted shifts for the \texorpdfstring{$^{13}$C}{13C} and \texorpdfstring{$^{15}$N}{15N} experimental benchmarks}

\begin{figure}[h!]
    \centering
    \includegraphics[width=1\linewidth]{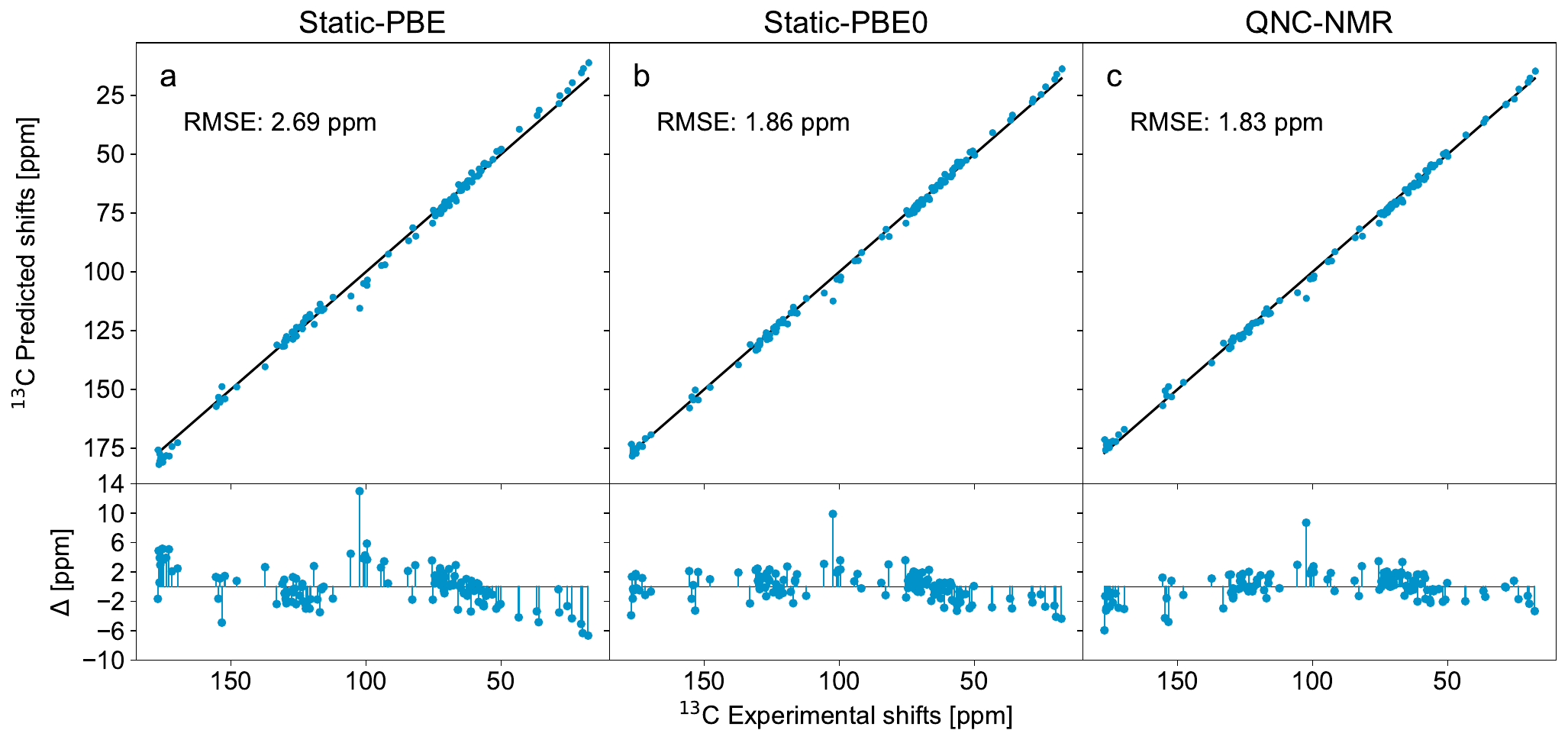}
    \caption{Comparison of experimental and predicted $^{13}$C chemical shifts computed with ShiftML3, using Static-PBE optimized geometries (a), Static-PBE0 optimized geometries (b), and the QNC-NMR protocol (c). The black diagonal line in each panel corresponds to a perfect correlation. The accuracy of the predictions is evaluated with the root-mean square error (RMSE). On the lower panels, $\Delta$ is the difference between the predicted and the experimental chemical shift ($\Delta$ = $\delta_{pred}$ – $\delta_{exp}$).}
    \label{fig:13C_bench_fix_slope}
\end{figure}

In \rev{Figure \ref{fig:13C_bench_fix_slope}} there are the parity plots for the $^{13}$C experimental benchmark from Ramos et al. \cite{ramosInterplayDensityFunctional2024} using the linear fit to map the experimental chemical shifts against calculated shieldings with a slope of  -1. As discussed in the main text, the Static-PBE0 geometries using the PET-MOLS led an improvement of the accuracies of the \rev{31}\% (2.6\rev{9} to 1.\rev{86} ppm) with respect to the Static-PBE geometries. An improvement of 32\% (2.6\rev{9} to 1.8\rev{4} ppm)  is further achieved with the new proposed QNC-NMR protocol.

\begin{figure}[h!]
    \centering
    \includegraphics[width=1\linewidth]{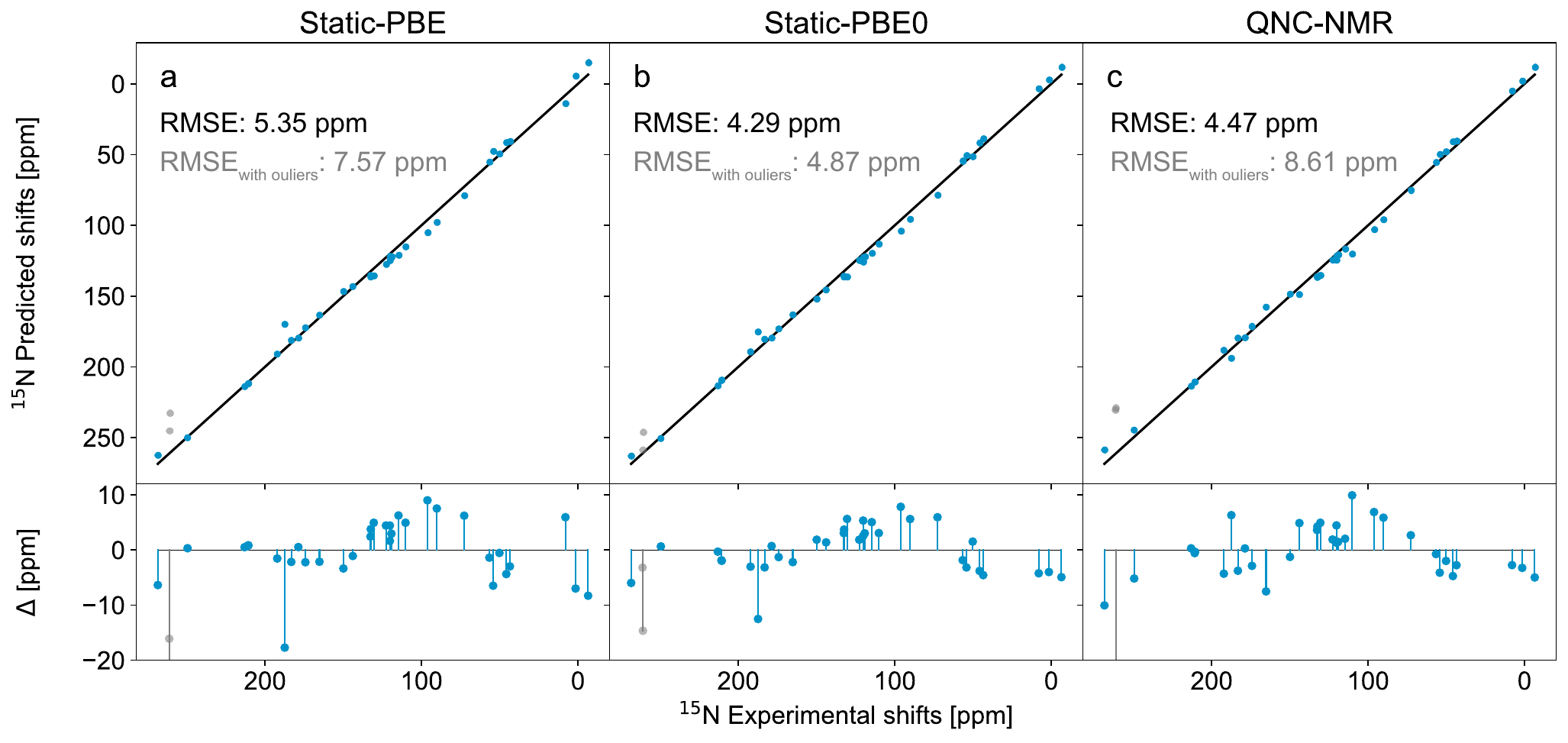}
    \caption{Comparison of experimental and predicted $^{15}$N chemical shifts computed with ShiftML3, using Static-PBE optimized geometries (a), Static-PBE0 optimized geometries (b), and the QNC-NMR protocol (c). The black diagonal line in each panel corresponds to a perfect correlation. The accuracy of the predictions is evaluated with the root-mean square error (RMSE) without (in black) and with the outliers (in grey). On the lower panels, $\Delta$ is the difference between the predicted and the experimental chemical shift ($\Delta$ = $\delta_{pred}$ – $\delta_{exp}$). Outlier local atomic environments not included in the fitting are displayed in color grey. Those correspond to the structures GEHHIL and GEHHEH whose Schiff base motifs have highly uncertain protonation state and are discussed in more detail the main text.}
    \label{fig:15N_bench_fix_slope}
\end{figure}

 In Figure \rev{\ref{fig:15N_bench_fix_slope}}, the$^{15}$N experimental benchmark are show for the three approaches. Prediction of Static-PBE0 optimized geometries and using the QNC-NMR protocol employing the PET-MOLS improves the predictions by \rev{2}0\% (\rev{4.29} ppm) and \rev{16}\% (\rev{4.47} ppm), respectively, with respect to the Static-PBE geometries (5.\rev{35} ppm).

\subsection{Linear Regression}
For the analysis, a  linear regression $\delta = a\sigma +b$ with a slope of a=-1 and  intercept (b) fit to give best agreement between predicted shielding ($\sigma$) and experimental chemical shift ($\delta$) were used. For completeness, we have included regression parameters where the slope is also optimized. These optimized parameters are listed in Table \ref{tab:RMSE_bench}. 

In Tables \ref{tab:1H_tables}-\ref{tab:table_15N} there is all data used for the mapping of the experimental shifts as a function of the predicted shieldings. All local atomic environments are considered for all three experimental benchmark, except the two $^{15}$N sites corresponding to the Schiff base sites. The reason of excluding those sites is the sensitivity to the protonation state encoded in the QNC-NMR protocol and it is further discussed in the main text.

\begin{figure}[h!]
    \centering
    \includegraphics[width=1\linewidth]{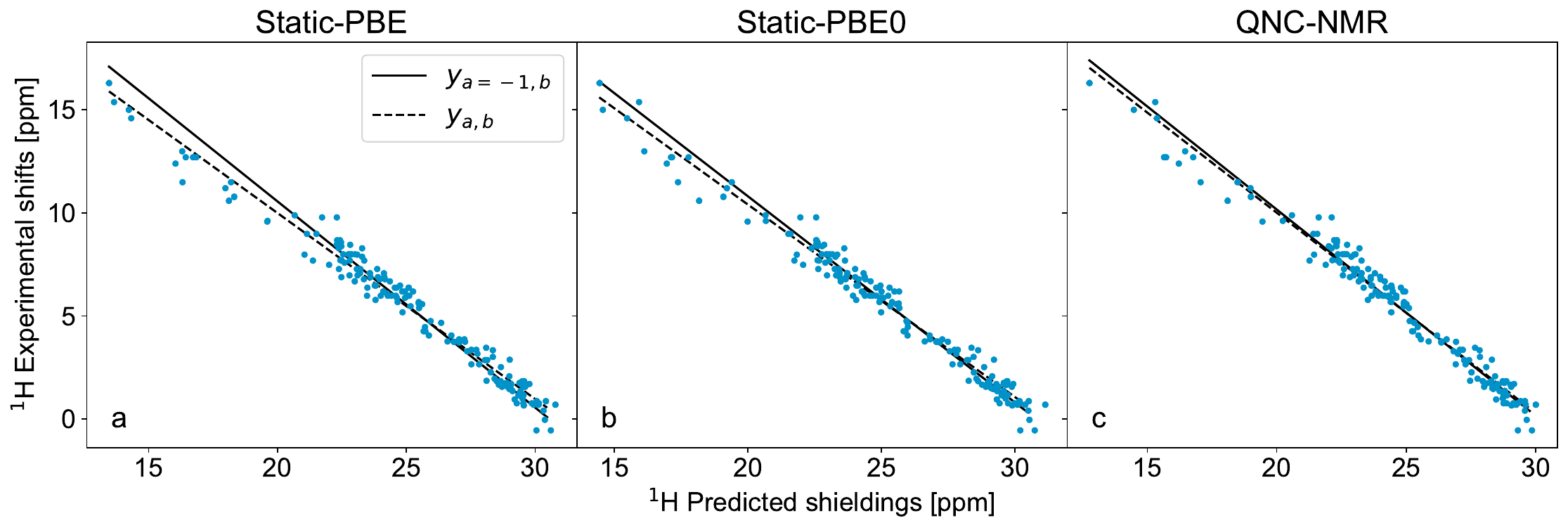}
    \caption{\rev{$^1$H} experimental chemical shifts as a function of the ShiftML3 predicted shieldings for the Static-PBE optimized geometries (a), Static-PBE0 optimized geometries (b), and the QNC-NMR protocol (c). In a black solid line ($\delta = a\sigma +b$) is the fitted linear regression keeping the slope fixed ($\delta_{a=-1, b}$, used in the main text); and in black dashed line ($\delta_{a,b}$) is the linear fit optimizing both the slope and intercept.}
    \label{fig:1H_slopes}
\end{figure}

\begin{figure}[h!]
    \centering
    \includegraphics[width=1\linewidth]{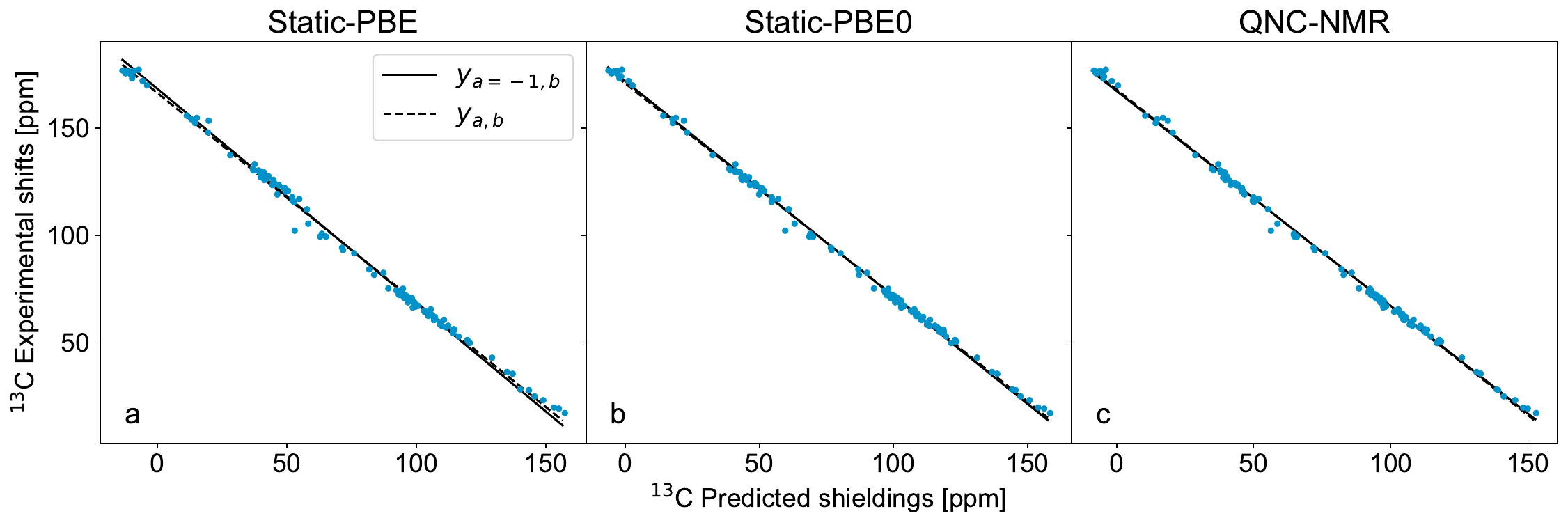}
    \caption{\rev{$^{13}$C} experimental chemical shifts as a function of the ShiftML3 predicted shieldings for the Static-PBE optimized geometries (a), Static-PBE0 optimized geometries (b), and the QNC-NMR protocol (c). In a black solid line ($\delta = a\sigma +b$) is the fitted linear regression keeping the slope fixed ($\delta_{a=-1, b}$, used in the main text); and in black dashed line ($\delta_{a,b}$) is the linear fit optimizing both the slope and intercept.}
    \label{fig:13C_slopes}
\end{figure}

\begin{figure}[h!]
    \centering
    \includegraphics[width=1\linewidth]{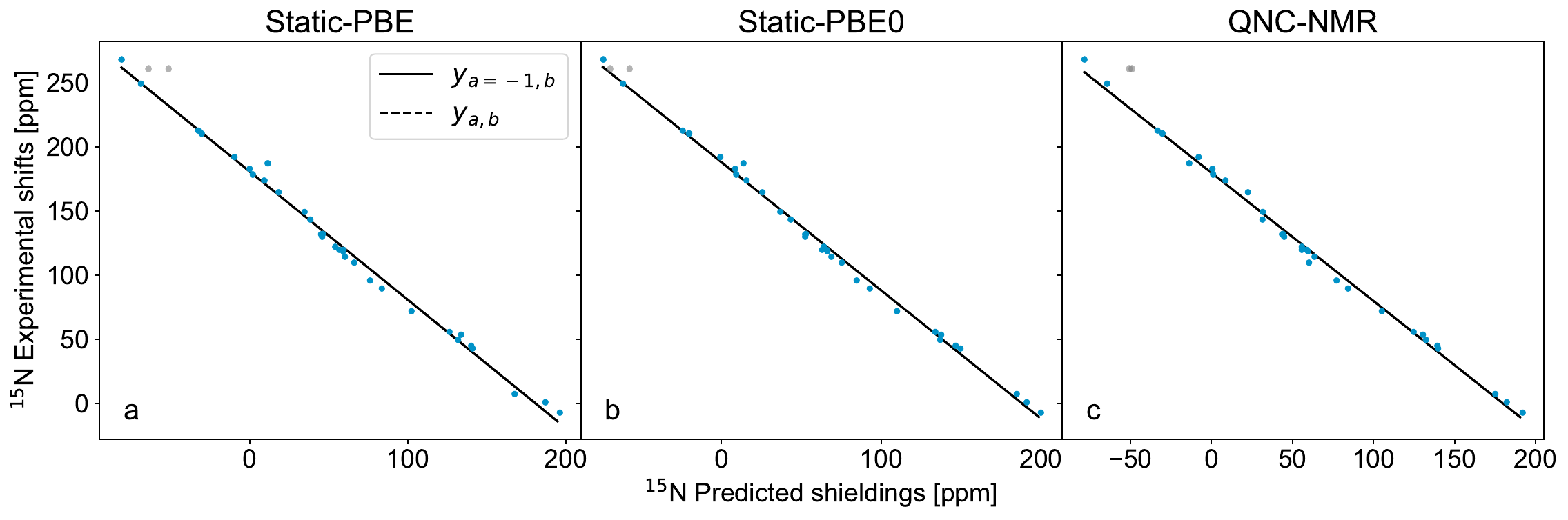}
    \caption{\rev{$^{15}$N} experimental chemical shifts as a function of the ShiftML3 predicted shieldings for the Static-PBE optimized geometries (a), Static-PBE0 optimized geometries (b), and the QNC-NMR protocol (c). In a black solid line ($\delta = a\sigma +b$) is the fitted linear regression keeping the slope fixed ($\delta_{a=-1, b}$, used in the main text); and in black dashed line ($\delta_{a,b}$) is the linear fit optimizing both the slope and intercept. Outlier local atomic environments not included in the fitting are displayed in color grey. Those correspond to the structures GEHHIL and GEHHEH whose Schiff base motifs have highly uncertain protonation state and are discussed in more detail the main text.}
    \label{fig:15N_slopes}
\end{figure}

\begin{table}[htbp!]
  \centering
  \caption{Linear fitting parameters (slope, $a$; and intercept, $b$) for the three different benchmarks ($^1$H, $^{13}$C and $^{15}$N) of crystalline structures using the Static-PBE geometries, the Static-PBE0 geometries, and the QNC-NMR protocol. Both approaches in the fitting — keeping the slope fixed to $-1$ and fitting the intercept, and fitting both the slope and intercept — are considered with their respective values and RMSE (root-mean square error).}
  \label{tab:RMSE_bench}
  \begin{tabular}{c l c c c c c}
    \toprule
    \multirow{2}{*}{\textbf{Nuclei}} & 
    \multirow{2}{*}{\textbf{Method}} & 
    \multicolumn{2}{c}{\textbf{Fix slope ($a=-1$)}} & 
    \multicolumn{3}{c}{\textbf{Fit slope}} \\
    \cmidrule(lr){3-4} \cmidrule(lr){5-7}
     & & \textbf{b} & \textbf{RMSE [ppm]} & \textbf{a} & \textbf{b} & \textbf{RMSE [ppm]} \\
    \midrule
    \multirow{3}{*}{\(\mathbf{^{1}H}\)} 
      & Static-PBE         & \rev{30.57}  & \rev{0.65} & \rev{-0.9036} & \rev{28.08} & \rev{0.53} \\
      & Static-PBE0   & \rev{30.81}  & \rev{0.56} & \rev{-0.9329} & \rev{29.06} & \rev{0.51} \\
      & QNC-NMR        & \rev{30.16}  & \rev{0.55} & \rev{-0.9705} & \rev{29.41} & \rev{0.54} \\
    \midrule
    \multirow{3}{*}{\(\mathbf{^{13}C}\)} 
      & Static-PBE         & \rev{168.23} & \rev{2.69} & \rev{-0.9737} & \rev{166.29} & \rev{2.44} \\
      & Static-PBE0   & \rev{171.78} & \rev{1.86} & \rev{-0.9865} & \rev{170.74} & \rev{1.76} \\
      & QNC-NMR        & \rev{167.20} & \rev{1.83} & \rev{-1.0097} & \rev{167.90} & \rev{1.80} \\
    \midrule
    \multirow{3}{*}{\(\mathbf{^{15}N}\)} 
      & Static-PBE    & \rev{180.85} & \rev{5.35} & \rev{-0.9992} & \rev{180.80} & \rev{5.35} \\
      & Static-PBE0   & \rev{188.13} & \rev{4.29} & \rev{-1.0000} & \rev{188.14} & \rev{4.29} \\
      & QNC-NMR       & \rev{179.92} & \rev{4.47} & \rev{-1.0010} & \rev{179.97} & \rev{4.47} \\
    \bottomrule
  \end{tabular}
\end{table}

\rev{\subsection{Computational cost for geometry optimization (DFT vs PET-MOLS)}

The overall computational cost (i.e. wall time) on a supercomputer cluster can be split in geometry optimization and shielding prediction. The cocaine polymorph, for example, using the PET-MOLS reduced the DFT geometry optimization step from 1 hour and 6 min to 22 seconds. In other words, the geometry optimization step is sped up ~200 times using the MLIP compared to traditional DFT.

}

\newpage
\pagebreak
\section{Amorphous drugs. Molecular dynamics and PIMD simulations} \label{sec:amorphous data}

\subsection{Glucagon-like peptide-1 receptor agonist (GLP-1RA)} \label{sec:GLP-1RA}

\begin{figure}[h!]
    \centering
    \includegraphics[width=1.0\linewidth]{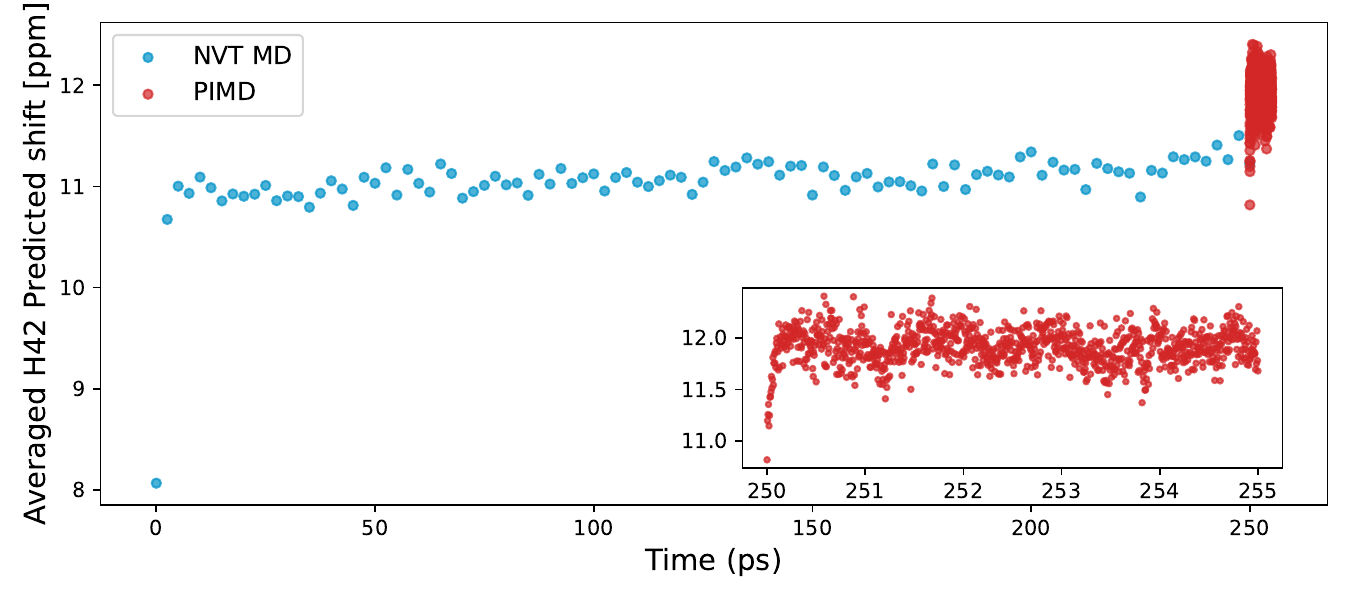}
    \caption{The averaged H42 predicted chemical shift \rev{of structure 1-Cl} through the NVT MD (in blue) and PIMD (in red) simulation timestep for one of the 17 MD frames. The inset is a zoomed-in region of averaged H42 predicted shift along the PIMD simulation.}
    \label{fig:MD_PIMD}
\end{figure}

\rev{The 17 selected MD frames were first pre-}equilibrated with  NVT MD for about 2\rev{5}0 ps at 300 K  and the subsequent \rev{5} ps PIGLET-accelerated-PIMD simulation using 8-beads, employing the PET-MOLS in both simulations. In Figure \rev{\ref{fig:MD_PIMD}}, we show the averaged acidic H42 chemical shifts along the trajectories for both simulations.

\pagebreak

\rev{\subsection{AZD4625} \label{sec:AZD4625}}

\begin{figure}[h!]
    \centering
    \includegraphics[width=0.85\linewidth]{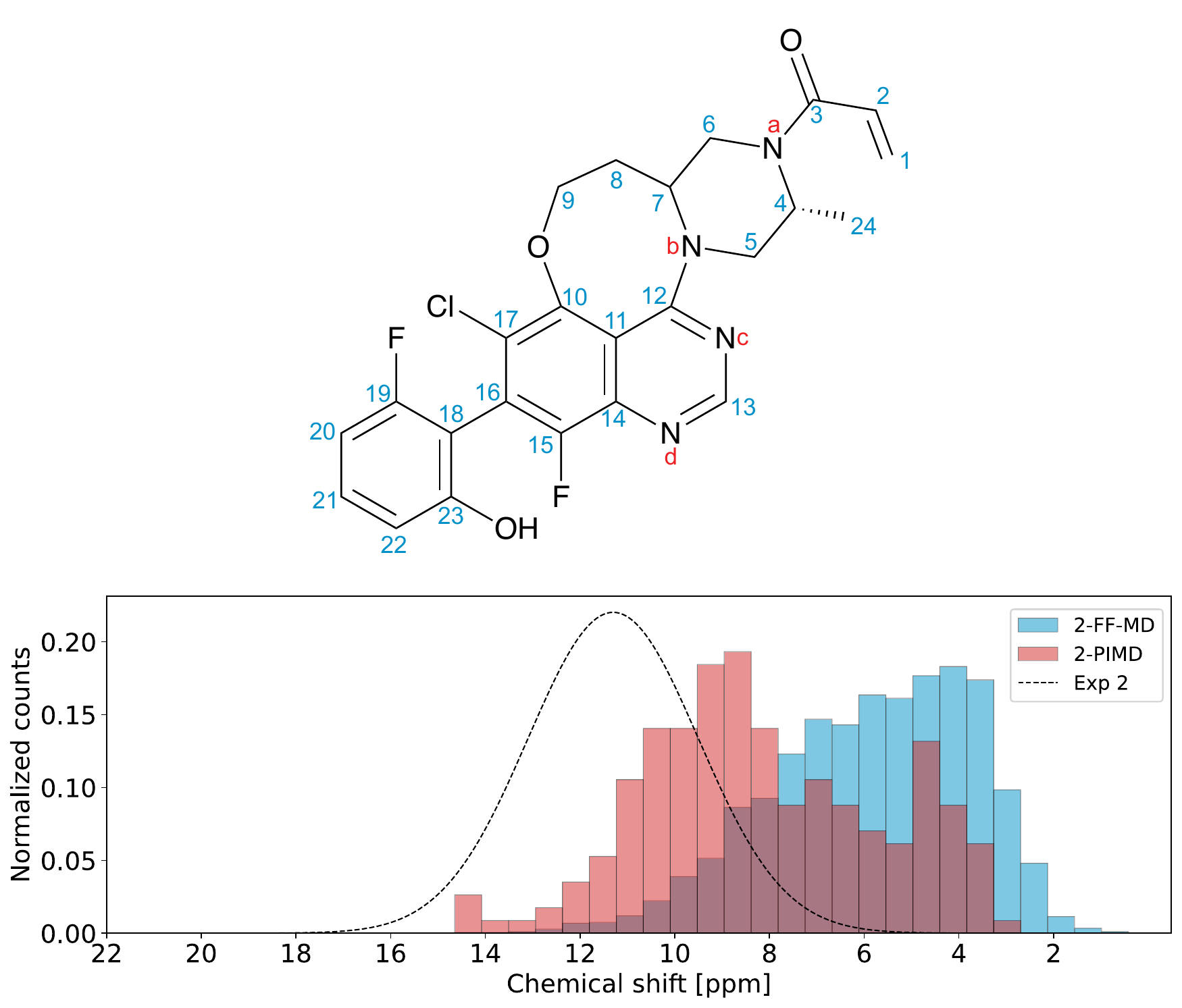}
    \caption{Predicted chemical shift distribution for all the molecular environments for structure 2 from 24 random snapshots from the FF-MD trajectories (blue), and the distribution of predicted shifts using the QNC-NMR protocol (red), compared to the experimentally determined chemical shift distribution (dashed black line) for the labile OH proton (H23 in the structure). The experimental distribution is taken from ref \cite{cordovaAtomiclevelStructureDetermination2023}}
    \label{fig:AZD4625_shifts}
\end{figure}

\rev{
The QNC-NMR protocol was also applied to another amorphous drugs, AZD4625 (named, structure 2), \cite{cordovaAtomiclevelStructureDetermination2023} whose labile predicted chemical shift distribution was off with respect to the experimentally observed one. 

That test case, was done in 24 randomly selected snapshots from the FF-MD trajectories form the original publication containing 128 molecules with 55 atoms each. \cite{cordovaAtomiclevelStructureDetermination2023} Those were pre-equilibrated for ~250 ps running a classical NVT MD at 300 K using employing the PET-MOLS followed by a 25 ps PIGLET-accelerated-PIMD simulation using 8-beads. All PIMD simulations of the experimental benchmark set were run at the respective experimental temperature. For the averaged chemical shift only the last 15 ps of the PIMD was considered.

In Figure \ref{fig:AZD4625_shifts} there is the predicted chemical shift distributions of the initial selected snapshots from the FF-MD trajectories and the averaged chemical shifts of those frames using the QNC-NMR protocol. We observe, again, the bimodal distribution as the GLP-1RA in the main text differentiating the hydrogen-bonded and non-hydrogen local atomic environments. Furthermore, the structure shows a notable improvements on the predicted chemical shift distributions for the hydrogen-bonded sites using the QNC-NMR approach that will help further structure selection (i.e. structures in best agreement with experimental data).

}

\pagebreak
\section{PET-MOLS training details}

\rev{In this study we introduce a new ML potential called \petpotential{} which we develop in two stages. First, a preliminary potential is constructed, covering elements HCNO. 
In a second stage the chemical space coverage of the potential is enlarged to 12 elements including S, F, P, Cl, Na, Ca,
Mg and K. In this section we list the training protocol of the potentials and the distribution of the final 12 element dataset that we generated to train \petpotential{}. Note that we separate the discussion of \petpotential{} development  into these two stages to demonstrate the improvement of simulated quantities using the 12 element version over the HCNO variant by the inclusion of additional training data, but also demonstrating that the simulation of molecular liquids, as observed in other MLIP studies \cite{zhaiShortBlanketDilemma2023} still remains challenging with modern MLIPs, even when adding additional training data. Unless stated otherwise, all presented evaluation results in the following studies were performed with the 12 element variant of \petpotential{}, as was used for all simulation results in the main text. }

\subsection{Machine learning interatomic potential}

We construct \rev{both the preliminary HCNO} \petpotential{} machine learning potential \rev{and the final 12-element parametrized varrian}t based on the point-edge transformer architecture (PET) from Pozdnyakov and Ceriotti \cite{NEURIPS2023_fb4a7e35}. Model hyperparameters were taken from a hyperparameter search from the PET-MAD project, a universal MLIP based on the PET architecture \cite{mazitovPETMADLightweightUniversal2025a}, that aimed to find optimal model hyperparameters balancing evaluation cost and prediction accuracy of PET models trained on large datasets (see also supplementary Figure 1 in Ref.~\cite{mazitovPETMADLightweightUniversal2025a}). Model hyperparameters were adopted without further optimization. The training dataset of PET-MAD contained also a portion of organic crystals, albeit computed at the PBEsol GGA DFT level of theory and we expect a good transferability of the PET-MAD hyperparameters to \petpotential{} trainings. \rev{The HCNO-} \petpotential{} was trained for 3,000 epochs in total.
\rev{The 12-elements \petpotential{} was trained starting directly from a PET-MAD initialized model checkpoint, increasing both the model stability and decreasing training times. Using rigid composition shifts, the energy scale of the PBE0-MBD data is automatically brought to the same energy scale of the PBEsol pretrained PET-MAD potential allowing for continued training. The 12-elements \petpotential{} was trained for 200 epochs.}

\rev{For the creation of the initial training set for training the HCNO-variant of \petpotential{}, the portion of structures containing only elements H,C,N,O from the ShiftML2/3 training and test databases were selected.} Originally these molecular compositions were selected via farthest point sampling (FPS) from the Cambridge Structural Database (CSD) to cover a broad range of structural motifs. The pool of candidates of the CSD-sampled database was limited to structures containing up to 200 atoms to restrict memory and computational costs of reference computation.
Experimental crystal structures were geometry relaxed using PBE-D2 and for each unique composition four thermally distorted structures were generated from short molecular dynamics simulations at 300~K using PBE-D2.
A total of 2,237 unique compositions was taken from the ShiftML2/3 dataset, forming together with the thermally distorted structures a dataset containing 11,176 structures. 
The dataset was split into a training, validation and testing set along compositions, assigning the relaxed and thermally distorted structures of one compositions exclusively to either set in order  to evaluate the capabilities of \rev{the HCNO-}\petpotential{} against unseen compositions and their distorted structures.
We train \rev{the HCNO-}\petpotential{} on a dataset of 2,013 unique organic crystal compositions with a total of 10,056 training structures. A validation set of 56 unique compositions with 280 structures is used for the learning rate scheduler during training. Prediction errors are evaluated on a test set containing 168 unique compositions with 840 structures. 
\rev{In order to train the 12 element variant of \petpotential{} we continued to add structures from the ShiftML2 training set containing additional elements, S, F, P, Cl, Na, Ca, Mg and K . Note that for the additional structures only one additional thermally distorted structure per composition was added, focusing on chemical diversity in the creation of the database. For the final 12 element dataset we choose to create new training, validation and test set that are strictly oriented at the ShiftML3 training,validation and testing set. This means that we ultimately have generated PBE0+MBD labels for subsets of the compositions of the ShiftML3 training, validation and test sets, as far as our computational budget permitted, with having additional thermally distorted structures for the HCNO space of the ShiftML3 selected training crystals. The benefit of this approach is a more honest assessment of both shielding-ML model and potential can be made, since there is no information leakage between the shielding and MLIP training and validation sets. Finally we have added 163 additional liquid water structures, of boxes with 64 water molecules to the training sets. These were sampled from NPT-PIMD simulations using the initial HCNO \petpotential{} and then also recomputed to obtain PBE0+MBD reference labels. Missing structures from the original ShiftML2 dataset, are mostly due to excessive memory requirements of hybrid functional computations, or instabilities of the MBD@rsSCS dispersion model when metal ions are present and in future work we will work on adding the missing structures, potentially switching to the MBD-NL dispersion model and changing the algorithm how Fock matrices are constructed. We train \petpotential{} with 12 element coverage on a dataset of 6,834 unique organic crystal compositions with a total of 22,483 training structures. A validation set of 206 unique compositions with 411 structures is used for the learning rate scheduler during training. Prediction errors are evaluated on a test set containing 510 unique compositions with 1,014 structures.  }

Training curves of \petpotential{} are shown in the SI~Fig.~\ref{fig:training_curve}. A more detailed analysis of the properties of \petpotential{} is discussed in the following sections. \rev{We also give an overview over the training data generated in the following two subsections.}

\subsection{\rev{Preliminary \petpotential{}}\,\,Dataset Summary}
\rev{ Table~\ref{tab:datasets_distr_hcno} lists the number of unique compositions, structures and total number of atoms in the training, validation and test sets used to train and evaluate the preliminary \petpotential{} with element-coverage HCNO.}

\begin{table}[ht]
    \centering
    \caption{Number of unique compositions $N_{\text{compositions}}$, number of structures $N_{\text{structures}}$ and total number of atoms $N_{\text{atoms}}$ in training, validation and test set used to train \rev{the prelimary HCNO-\petpotential{}.  }}
    \label{tab:datasets_distr_hcno}
    \resizebox{0.55\textwidth}{!}
    {\begin{tabular}{lccc}
        \toprule
        Set & $N_{\text{compositions}}$ & $N_{\text{structures}}$ & $N_{\text{atoms}}$ \\
        \midrule
        Training     & 2,013 & 10,056 & 900,568 \\
        Validation &  56  & 280 & 26,130 \\
        Test & 168 & 840 & 71,145 \\
        \bottomrule
    \end{tabular}}
\end{table}

\subsection{\rev{12-element \petpotential{}}\,\,Dataset Summary}
\begin{figure}[h!]
    \centering
    \includegraphics[width=0.9\linewidth]{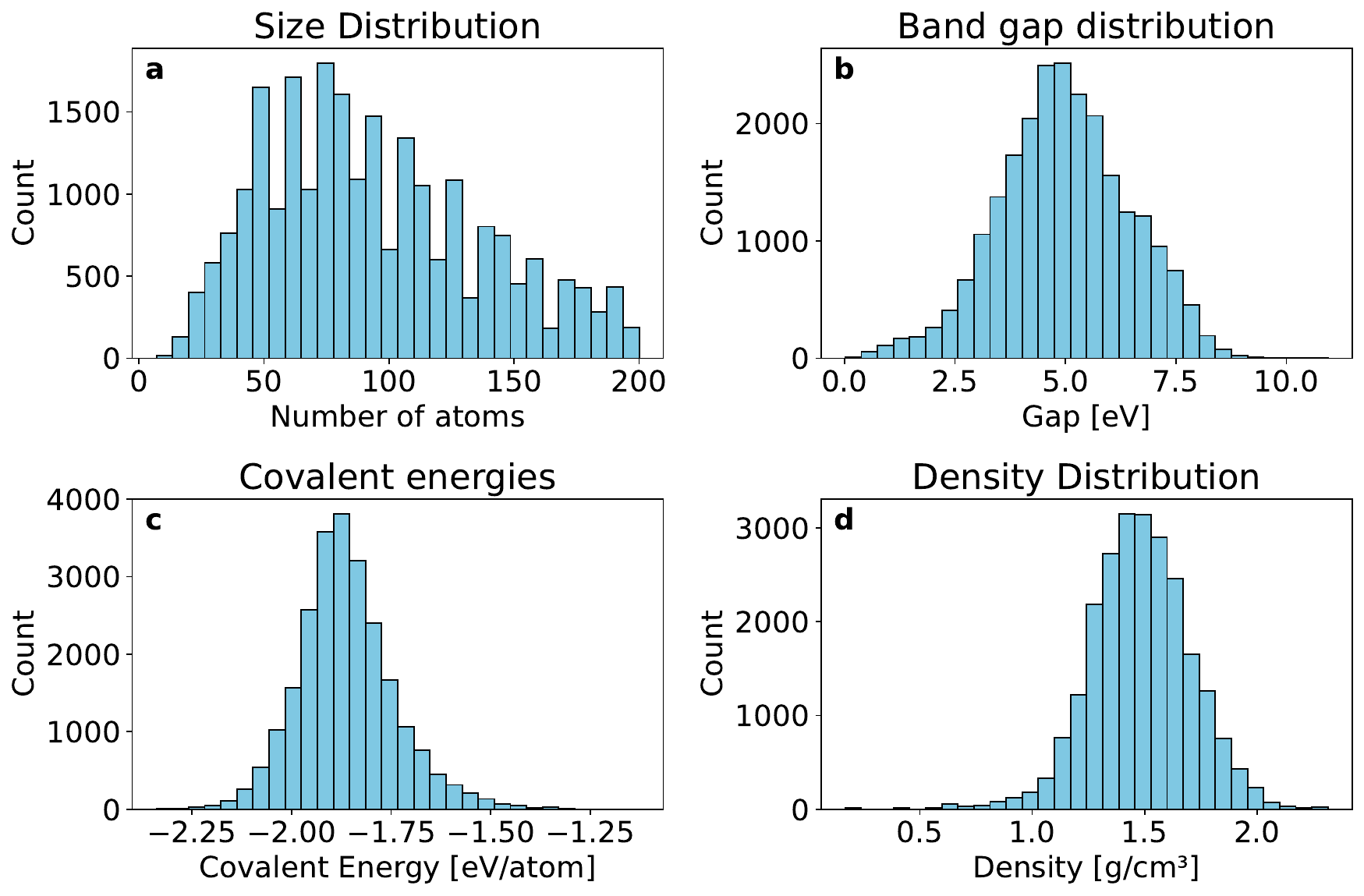}
    \caption{Distribution of properties of the joint training, validation and test sets. Panel (a) shows the distribution of atoms per structure, panel (b) shows the distribution of  electronic band gaps, panel (c) the distribution of covalent energies and panel (d) shows the distribution of densities of the joint datasets.}
    \label{fig:distribution_in_data}
\end{figure}

\rev{After the construction of an initial potential parametrized for HCNO-materials, we add additional structures.
In Figure~\ref{fig:distribution_in_data} we show the distribution of selected properties of the PBE0+MBD reference dataset computations that we introduce in this work and use to train \petpotential{} with 12 element coverage.
}


\begin{table}[ht]
    \centering
    \caption{\rev{Number of unique compositions $N_{\text{compositions}}$, number of structures $N_{\text{structures}}$ and total number of atoms $N_{\text{atoms}}$ in training, validation and test set used to train \petpotential{} for all 12-elements.  }}
    \label{tab:datasets_distr}
    \resizebox{0.55\textwidth}{!}
    {\begin{tabular}{lccc}
        \toprule
        Set & $N_{\text{compositions}}$ & $N_{\text{structures}}$ & $N_{\text{atoms}}$ \\
        \midrule
        Training     & 6,834 & 22,483 & 2,073,679  \\
        Validation &  206  & 411 & 46,196 \\
        Test & 510  & 1,014 & 116,559 \\
        \bottomrule
    \end{tabular}}
\end{table}

\pagebreak
\section{Evaluation of \petpotential{}}

We evaluate \petpotential{} predictions of the covalent energy $E_{\text{cov}}$ on a test set containing \rev{1,014} structures of \rev{510} unique compositions.
The covalent energy is  defined by removing isolated atom energies $\varepsilon^{\text{ref}}$ from the potential DFT/MLIP predicted energies $V_{\text{pot}}$:

\begin{equation}
    E_{\text{covalent}} = V_{\text{pot}} - \sum_{j\in \text{Elem}_{\text{train}}}n_j\varepsilon_{j}^{\text{ref}}
\end{equation}

\begin{figure}[h!]
    \centering
    \includegraphics[width=\linewidth]{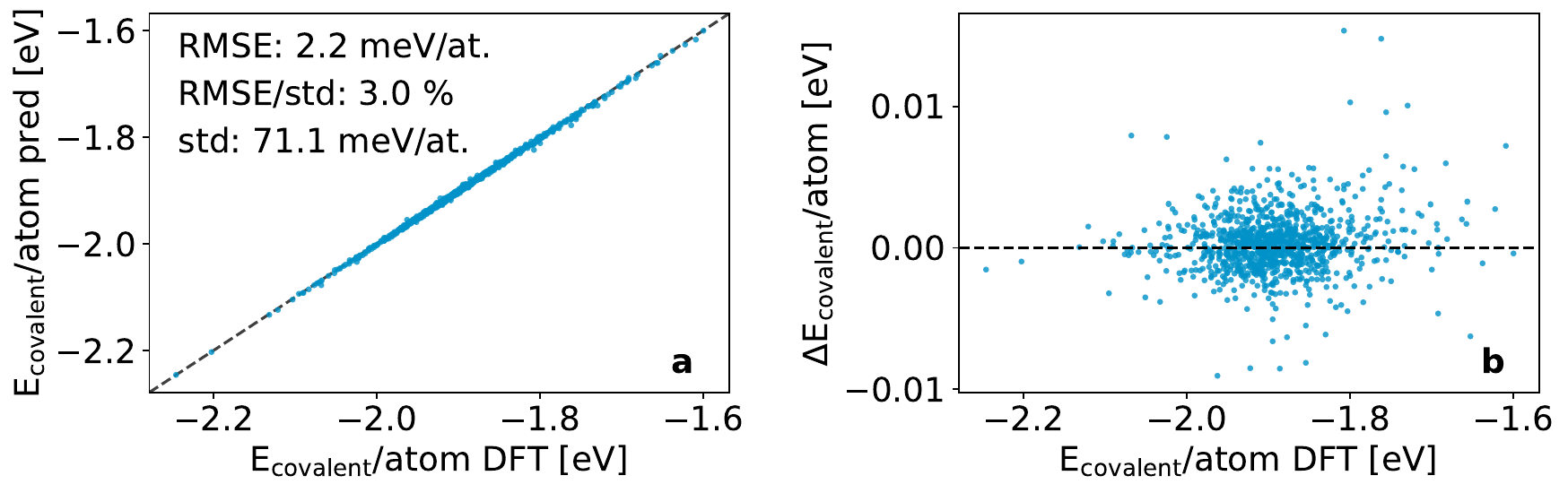}
    \caption{Parity plots of DFT covalent energy against predicted covalent energies by \petpotential{} (a) and DFT covalent energies against covalent energy prediction errors between \petpotential{} and DFT. Note the different x and y scales.}
    \label{fig:si_parity_energies}
\end{figure}

Note that composition energies are removed by a least squares fitted composition model in \petpotential{} and the total prediction errors on total potential energies $V_{\text{pot}}$ and $E_{\text{cov}}$ of \petpotential{} are identical modulo floating point errors, but the standard deviations of $V_{\text{pot}}$ would be artificially enlarged given that isolated atom energies can be significantly larger than covalent energies making the normalized RMSE error seem artificially low when they are not removed. Fig.~\ref{fig:si_parity_energies} (a) shows parity plots of predicted and computed covalent energies. Panel (b) shows the prediction errors of covalent energies( $\Delta E_{\text{covalent}}$ ) against the range of total covalent energies, highlighting that there are no biases in prediction errors towards particularly small or large covalent energies. 

\begin{figure}[h!]
    \centering
    \includegraphics[width=\linewidth]{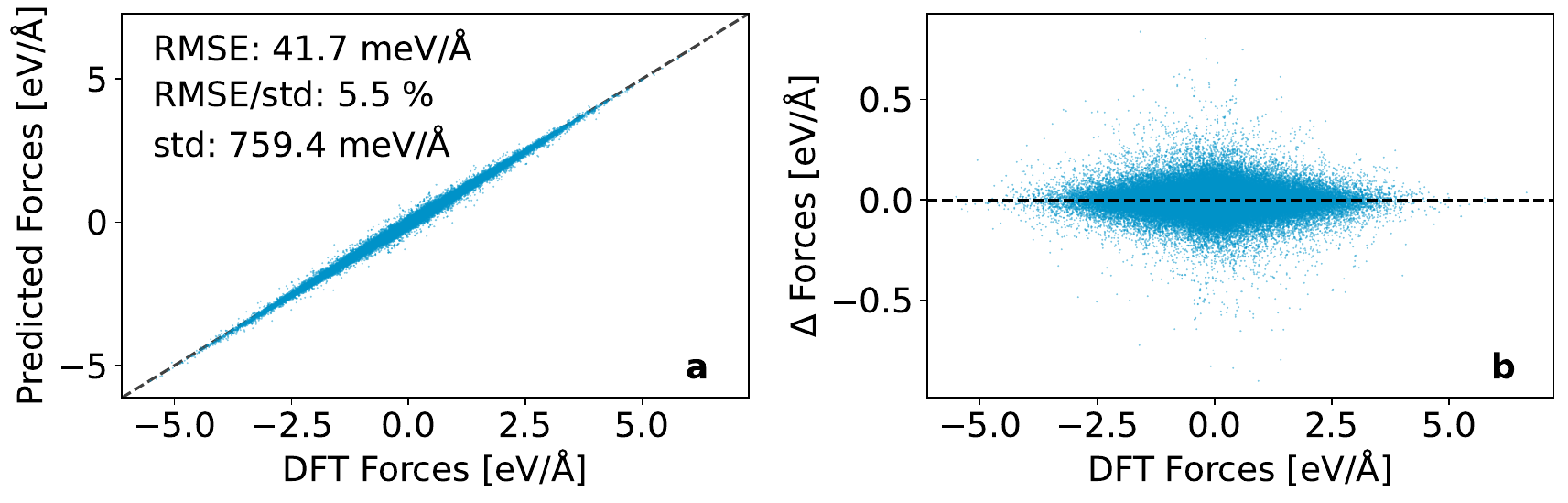}
    \caption{Parity plots of DFT force components against predicted forces by \petpotential{} (a) and force prediction errors between \petpotential{} and DFT against DFT forces. }
    \label{fig:si_parity_forces}
\end{figure}

We also evaluate the force prediction errors of \petpotential{} against DFT prediction errors and show parity plots of the \petpotential{} against DFT forces in Fig.~\ref{fig:si_parity_forces}.

\section{Selected atomistic modeling examples with \petpotential{}}
We evaluate \petpotential{} on selected modeling problems in condensed organic matter, that profit from both incorporation of quantum nuclear effects as well as first principle quality models of the potential energy surface. We used \petpotential{} to compute the density of liquid methane at high pressure, the equation of state (EOS) and average molecular volume of a benzene crystal Form 1, the radial distribution functions of liquid water, the temperature dependency of proton chemical shieldings in the organic acid-base pair SEDJUI and finally we use the chemical shielding distribution of acid protons in succinic acid to test the quantitative agreement of \petpotential{} with bespoke models built for succinic acid.

\subsection{Density of liquid methane}
\begin{figure}[h!]
    \centering
    \includegraphics[width=0.6\linewidth]{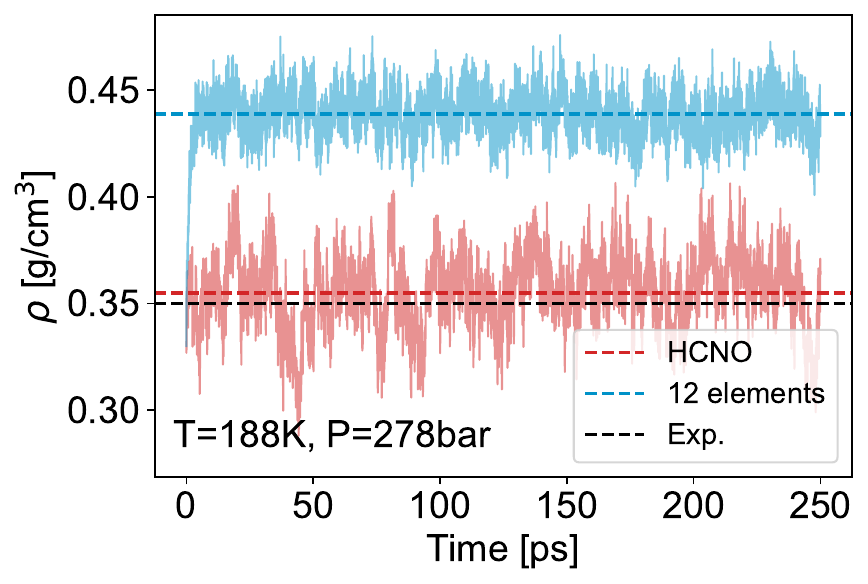}
    \caption{Path integral molecular dynamics simulations of liquid methane using PET-MOLS at 188~K at 278~bar. We show the time-evolution of the box density of the PIMD simulation.}
    \label{fig:timeseries_density}
\end{figure}

We compute the density of liquid methane with PIMD simulations with \petpotential{}. Veit and coworkers computed the equation of state of liquid methane using bespoke gaussian approximation potentials (GAP)\cite{veitEquationStateFluid2019}, revealing a strong sensitivity of the bulk density towards the parametrization of the van der waals interactions. Additionally, Pereyaslavets and coworkers found a non-negligible sensitivity of the bulk density of short alkanes towards nuclear quantum effects. \cite{pereyaslavetsImportanceAccountingNuclear2018} Veit and coworkers computed bulk densities with PIMD to account for such effects. We choose to adopt Veit's and Coworkers simulation parameters, simulating boxes of 100 methane molecules using colored Noise GLE barostats with coupling times of 100~fs and PIGLET thermostat with 16~beads, employing a timestep of 0.5~fs.\cite{ceriottiEfficientFirstPrinciplesCalculation2012,ceriottiNuclearQuantumEffects2009} Periodic systems are first equilibrated with short simulations in the NVT ensemble and then production simulations are run in the NPT ensemble. We simulate the methane system for a total of 250~ps. We find that the average density of \rev{HCNO}-\petpotential{} methane at 188~K and 278~bar is $0.355\pm 0.002$ $g/cm^{-3}$ which is in excellent agreement with the experimental values found to be 0.350 $g/cm^{-3}$ at the identical conditions \cite{goodwin1972densities}. Sampling uncertainties were computed by blocking analysis splitting the trajectory into 10 blocks. In Fig~\ref{fig:timeseries_density} we show the time-evolution of the density of the simulation box.
\rev{Repeating the same simulations with the 12 element PET-MOLS generates liquid methane densities of $0.439\pm 0.001$ $g/cm^{-3}$, deviating from the experimental reference. As stated earlier, simulating molecular liquids with MLIPs is challenging and has been recognized in earlier work\cite{zhaiShortBlanketDilemma2023}. Whether this can be attributed to the training protocol (finetuning from the PET-MAD checkpoint), the compressibility of CH4 being generally sensitive to MLIP prediction errors or distribution shift in the training data will be part of future work.}

\pagebreak

\subsection{\rev{Evolution of GEHHIL protonation free energy surface}}

\begin{figure}[h!]
    \centering
    \includegraphics[width=0.8\linewidth]{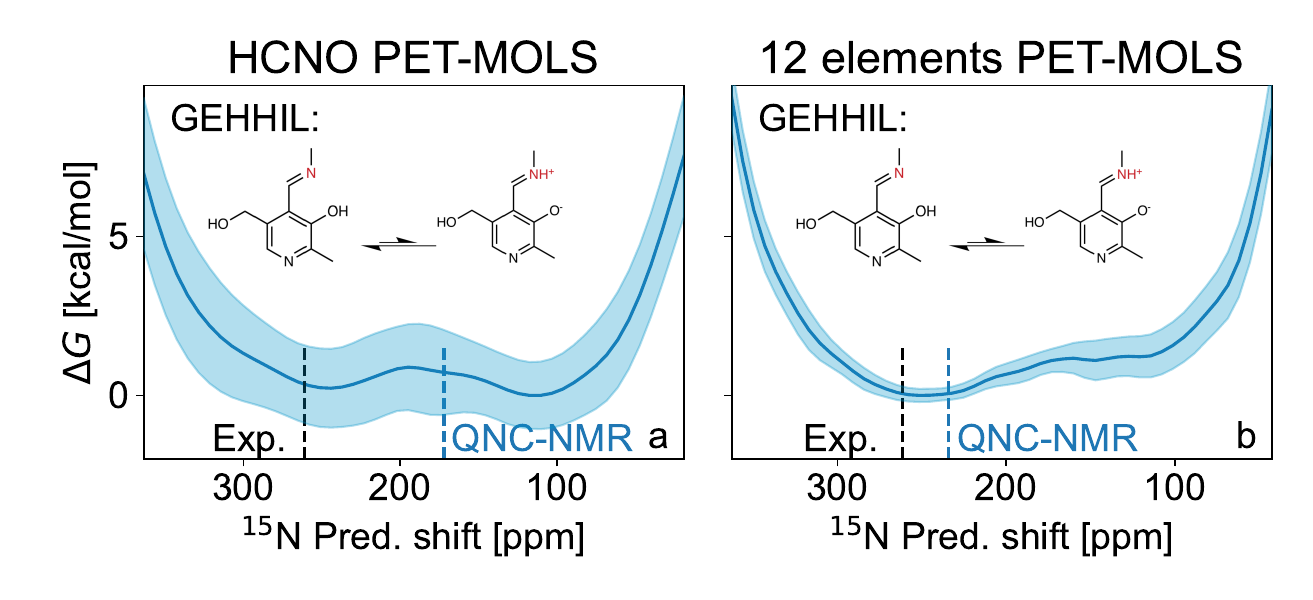}
    \caption{\rev{Comparing the free energy landscapes of GEHHIL protonation from the HCNO-\petpotential{} (a) and 12 elements-\petpotential{} (b). 12-element PET-MOLS predicted shift averages are in better agreement with experiment}}
    \label{fig:FES_compare}
\end{figure}

\rev{Comparing the free energy surface (FES) of the pathological GEHHIL protonation states using the HNCO-\petpotential{} and 12-elements \petpotential{} indicates a significant improvement of \petpotential{} after inclusion of additional training data. In Figure~\ref{fig:FES_compare} we compare the FES of GEHHIL of both potentials, indicating that the 12 element \petpotential{} predicts more accurate $^{15}$N chemical shifts of the Schiffs base nitrogen, whilst also making smaller uncertainty estimates over the HCNO potential. }

\subsection{Equation of state of the benzene crystal}
We compute the equation of state (EOS) of the benzene crystal using \petpotential{} and compute the average molecular volume of deuterated benzene C6D6 form I crystals in order to compare them with neutron scattering experiments of C6D6 performed at 123~K.\cite{jeffrey1987crystal} The EOS values are obtained from Murnaghan equation of state fits in order to compare them with computational values from \cite{ehlertR2SCAND4DispersionCorrected2021, zenFastAccurateQuantum2018}. In order to compute finite temperature average volumes of C6D6 we performed PIMD-NPT simulations with \petpotential{} at 123~K and 1~Bar using 16~Beads for 100~ps, employing a timestep of 0.5~fs , a PIGLET thermostat, and  a colored noise barostat and isotropic cell for barostating. We find that the average molecular volumes of C6D6 computed from PIMD simulation amounts to \rev{120.8~\AA{}$^3$} per molecule which is in good agreement with the experimental values of 118.3~\AA{}$^3$ per molecule \cite{jeffrey1987crystal}.

\begin{figure}[h!]
    \centering
    \includegraphics[width=\linewidth]{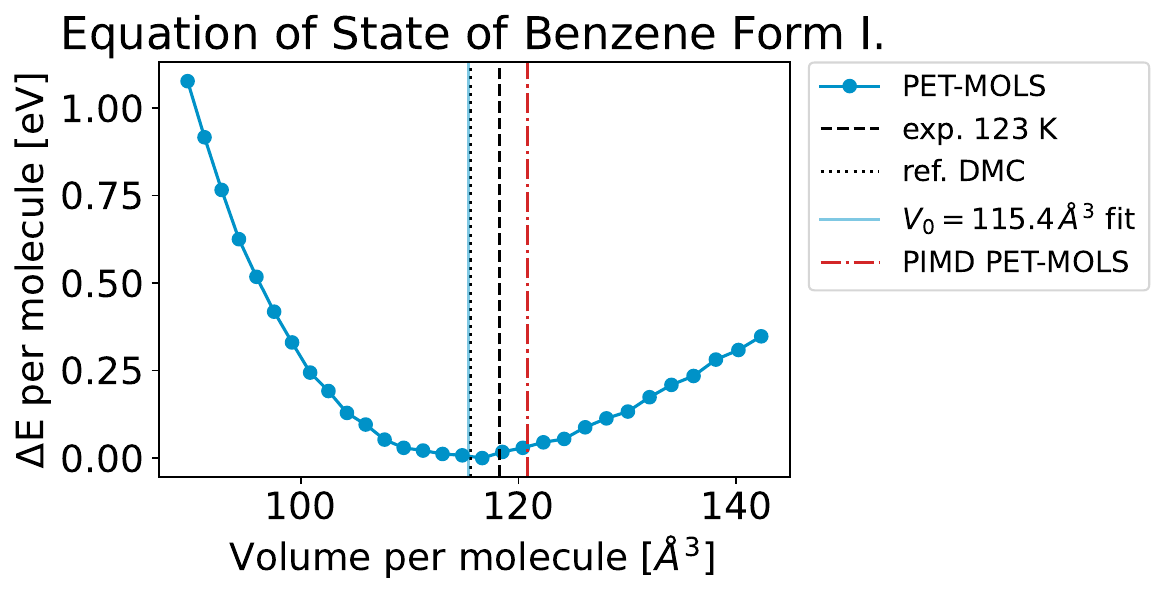}
    \caption{Equation of state (EOS) of benzene form I using  \petpotential{}. Additional molecular volumes of C6D6 PIMD averages at 1~bar and 123~K are shown for comparison, as well as experimental neutron scattering values at 123K \cite{jeffrey1987crystal} are indicated and static values of a diffusion quantum Monte Carlo (DMC) computation \cite{zenFastAccurateQuantum2018}.}
    \label{fig:placeholder}
\end{figure}
\newpage
\subsection{Radial distribution function of liquid water} 
\begin{figure}[h!]
    \centering
    \includegraphics[width=\linewidth]{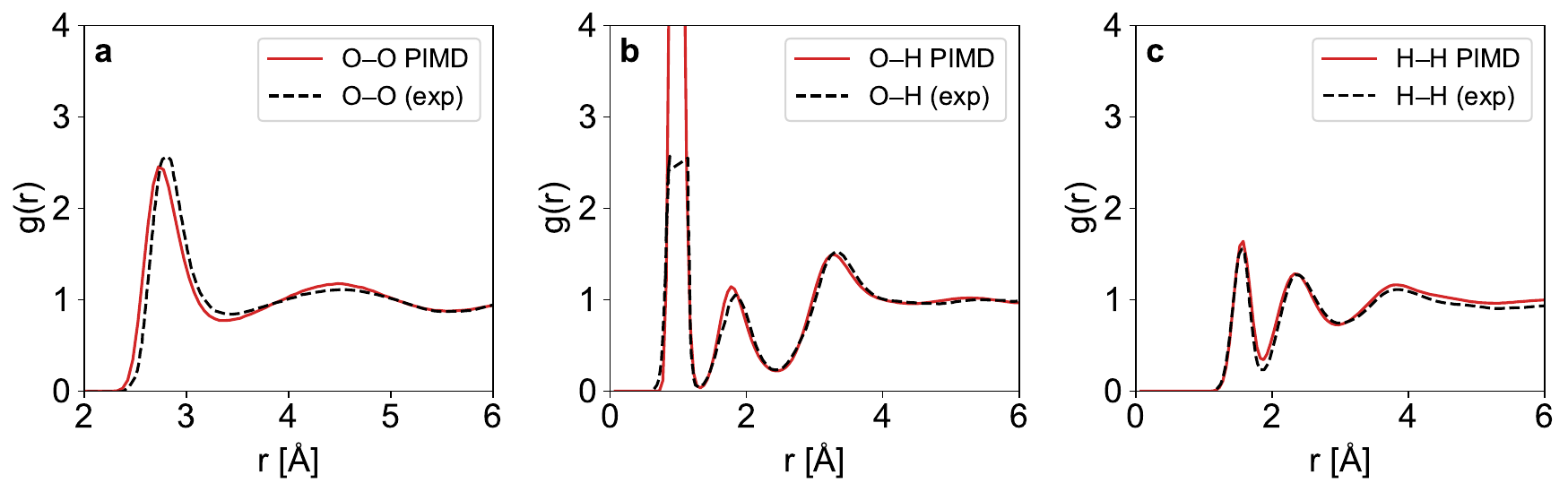}
    \caption{(a) Oxygen-Oxygen RDF , (b) Hydrogen-Oxygen RDF  and (c) Hydrogen-Hydrogen RDF of liquid water PIMD simulations with \petpotential{} at 390 K.}
    \label{fig:liquid_water_rdf}
\end{figure}

We compute radial distribution functions g(r) of liquid water using PIMD simulations and \petpotential{}. It is well known, that simulating radial distribution functions of liquid water require both an accurate description of the inter- and intramolecular interactions \cite{zhaiShortBlanketDilemma2023,reddyAccuracyMBpolManybody2016, monterodehijesDensityIsobarWater2024a} as well as the inclusion of NQEs \cite{paesaniAccurateSimpleQuantum2006, habershonCompetingQuantumEffects2009, chengInitioThermodynamicsLiquid2019b}. We note that in the construction of \petpotential{}, structures containing only water were not included in the training set. Water within the training set is present in the crystalline hydrates. We compute the element-wise radial distribution functions from NPT-PIMD simulations of boxes of 256 \ce{H2O} molecules at 390~K, compensating for the overestimated melting point of the \petpotential{} and compare them with experimentally measured radial distribution functions at 298~K.\cite{soperRadialDistributionFunctions2000b} We perform molecular dynamics using a PIGLET thermostat with 6-beads for 250~ps and 1~bar external pressure, employing a timestep of 0.5~fs. Even though the study of liquid water constitutes a clear out-of-distribution example for \petpotential{} we find that the radial distribution functions of the \petpotential{} water model are in acceptable agreement with the experimental radial distribution functions at 300~K. In Figure~\ref{fig:liquid_water_rdf} we compare computed and experimental radial distribution functions.

\begin{figure}[h!]
    \centering
    \includegraphics[width=\linewidth]{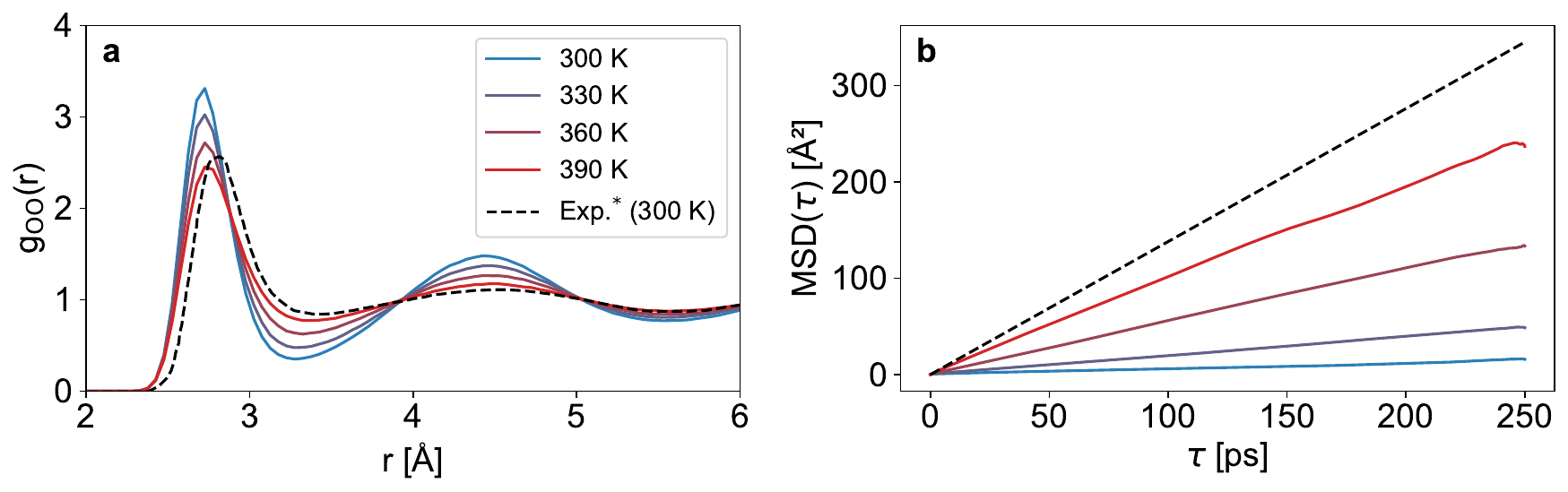}
    \caption{ Oxygen-Oxygen RDF of PIMD simulations of \petpotential{} for variable temperatures (a), and mean squared displacements (MSD) of oxygen of the same simulations (b). Note that the experimental  oxygen MSD values at 300 K are computed  from experimental diffusion coefficients. Computed diffusion coefficients have not been corrected for finite size effects and are influenced by the coloured-noise thermostat and should be regarded as a crude measure for the viscosity of the \petpotential{} water model at varying temperatures. }
    \label{fig:rdf_msd_liquid_H2O}
\end{figure}

 In Figure~\ref{fig:rdf_msd_liquid_H2O} we compare RDF and mean squared displacements (MSD) of Oxygen for liquid water at variable simulation  temperatures. From 390K to 360K, we observed a significantly lower MSD  and decrease in oxygen-oxygen RDF values between coordination shells, confirms that below 390~K the water must be considered undercooled.

\pagebreak

\subsection{Temperature dependency of \texorpdfstring{$^{1}$H}{1H} chemical shieldings in short hydrogen bonded base-acid pairs}

\begin{figure}[h!]
    \centering
    \includegraphics[width=0.7\linewidth]{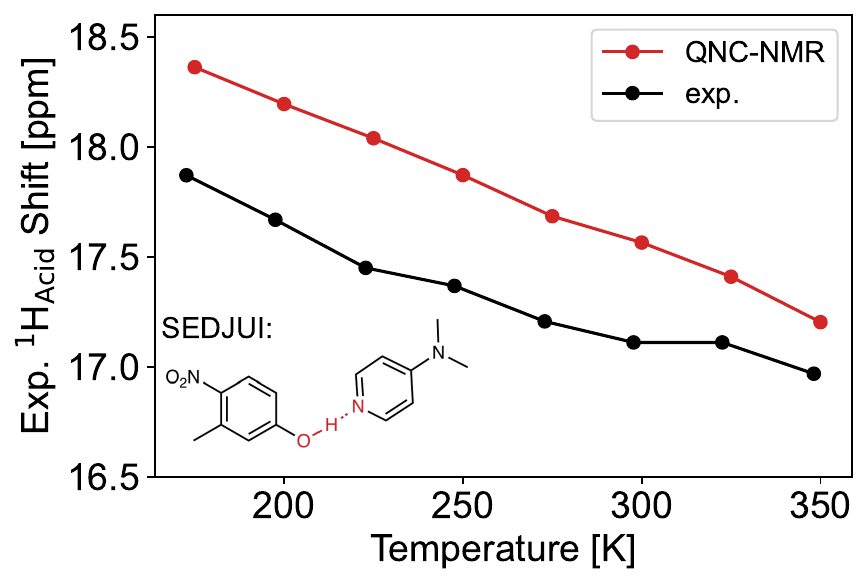}
    \caption{ Temperature dependency of the $^1$H chemical shifts of acidic protons of the SEDJUI cocrystal system, computed from PIMD simulations and experimental reference values from Ref.\cite{stocekImportanceNuclearQuantum2022a}. }
    \label{fig:temperature_dependence}
\end{figure}

We compute the temperature-dependency of $^1$H chemical shifts in the organic co-crystal system SEDJUI. \v{S}to\v{c}ek and coworkers identified a particular sensitivity of acidic $^{1}$H shifts in molecular crystals with short hydrogen bonds to quantum nuclear effects.\cite{stocekImportanceNuclearQuantum2022a} In the system SEDJUI (3-methyl-4-nitrophenol and 4-dimethylaminopyridine) the finite temperature effects and NQEs affect the fractional protonation states strongly and over a temperature range of 173~K to \rev{348}~K a conversion of salt to cocrystal is observed, as confirmed by $^1$H NMR shifts of acidic protons. We probe the temperature sensitivity of acid $^1$H shifts in SEDJUI models and run PIMD dynamics with \petpotential{} of SEDJUI structures. We run molecular dynamics for 150~ps at each temperature in the NPT ensemble at 1~bar pressure with an isotropic cell. We use a PILE-L thermostat for thermostating and a coupling time of 100~fs, as well as a langevin barostat with a coupling time of 100~fs. All simulations use 32 replicas and a timestep of 0.5~fs. We compute chemical shielding averages from the simulation trajectories with ShiftML3. Computed chemical shielding averages are converted to chemical shifts using a fixed regression slope of $a=-1$ and intercept of \rev{$b=30.16$}. In Figure~\ref{fig:temperature_dependence} we show the temperature dependency of both PIMD simulated chemical shifts and experimentally measured values. Note that unlike in the main text, in which we compare only relative chemical shielding differences $\Delta\sigma_{\text{iso}}$ to the chemical shifts at 173~K, we visualize in this additional figure the total chemical shift (after regression with fix slope and QNC-NMR offset), highlighting that the temperature dependency is captured exactly, whilst the absolute chemical shift value deviates by about 0.5~ppm which is in line with the prediction accuracy of $^1$H chemical shifts by QNC-NMR against experimental benchmark structures. This deviation can most probably be attributed to shortcomings in the reference GIPAW DFT calculations in the ShiftML3 training database.

\pagebreak

\subsection{\rev{Highlighting the importance of PIMD sampling, on the example of \texorpdfstring{$^{1}$H}{1H} chemical shieldings in short hydrogen bonded base-acid pairs}}

\begin{figure}[h!]
    \centering
    \includegraphics[width=0.95\linewidth]{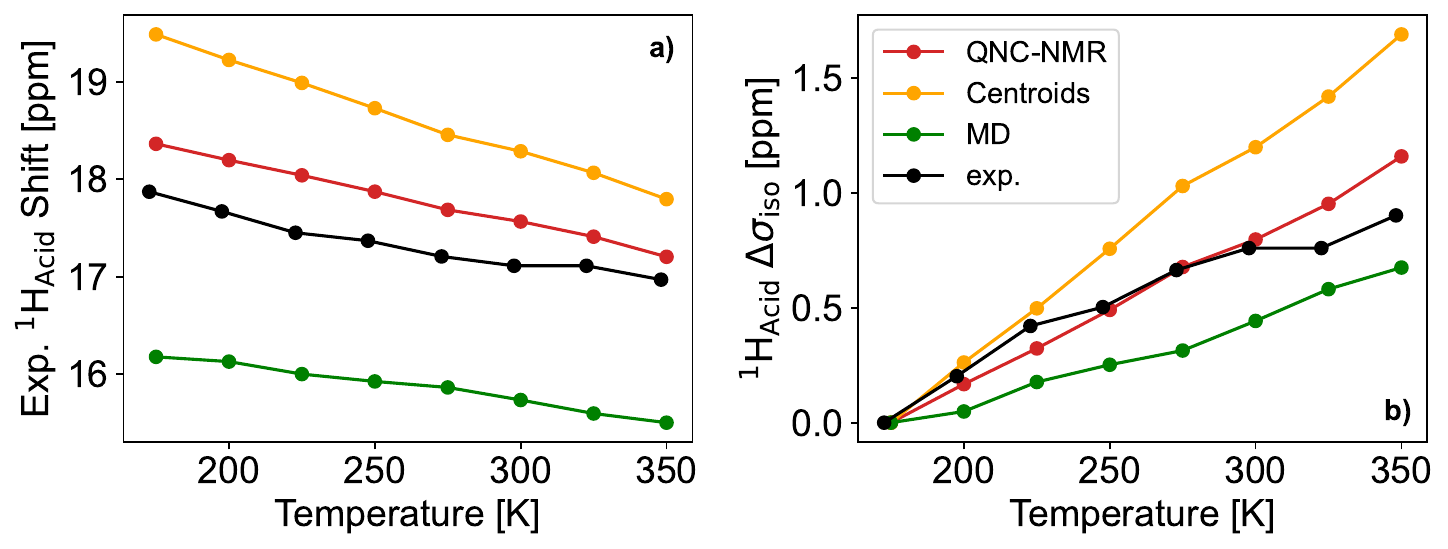}
    \caption{\rev{ a) Temperature dependency of the $^1$H chemical shifts of acidic protons of the SEDJUI cocrystal system, computed from PIMD simulations, BOMD simulations and PIMD average structures (centroids) and experimental reference values from Ref.\cite{stocekImportanceNuclearQuantum2022a}. Panel b) additionally shows the temperature dependency $\Delta\sigma_{\text{}iso}$ of the labile proton shielding relative to the value at 173~K. }}
    \label{fig:temperature_dependence_combined}
\end{figure}

\rev{We use the temperature dependency of the acidic proton in SEDJUI to demonstrate the effect of quantum nuclear effects and the explicit treatment of them with PIMD simulations in the QNC-NMR protocol. In Figure~ \ref{fig:temperature_dependence_combined} we compare the temperature evolution of the acidic chemical shift in SEDJUI computed using PIMD averages, averages computed from PIMD geometry averages (centroids) and Born-Oppenheimer MD simulations.  All simulations were performed with \petpotential{} and ShiftML3, any differences between the simulations are arising from the different treatments of quantum nuclear effect in the simulation protocols. The strong deviations between centroid and PIMD (QNC-NMR) chemical shift averages indicate a strong anharmonicity and delocalization of the ring polymer. This can be perhaps best be understood by describing the acid-base proton-transfer in SEDJUI with a double well potential, separating salt and cocrystal state. Conversely, chemical shift averages computed from Born Oppenheimer molecular dynamics simulations strongly underestimate the absolute chemical shift of the acidic proton by 2~ppm, furthermore underestimating finite temperature effects $\Delta\sigma_{\text{iso}}$.}

\subsection{\rev{Detailed discussion of uncertainty contributions in the QNC-NMR workflow}}

\begin{figure}[h!]
    \centering
    \includegraphics[width=1.00\linewidth]{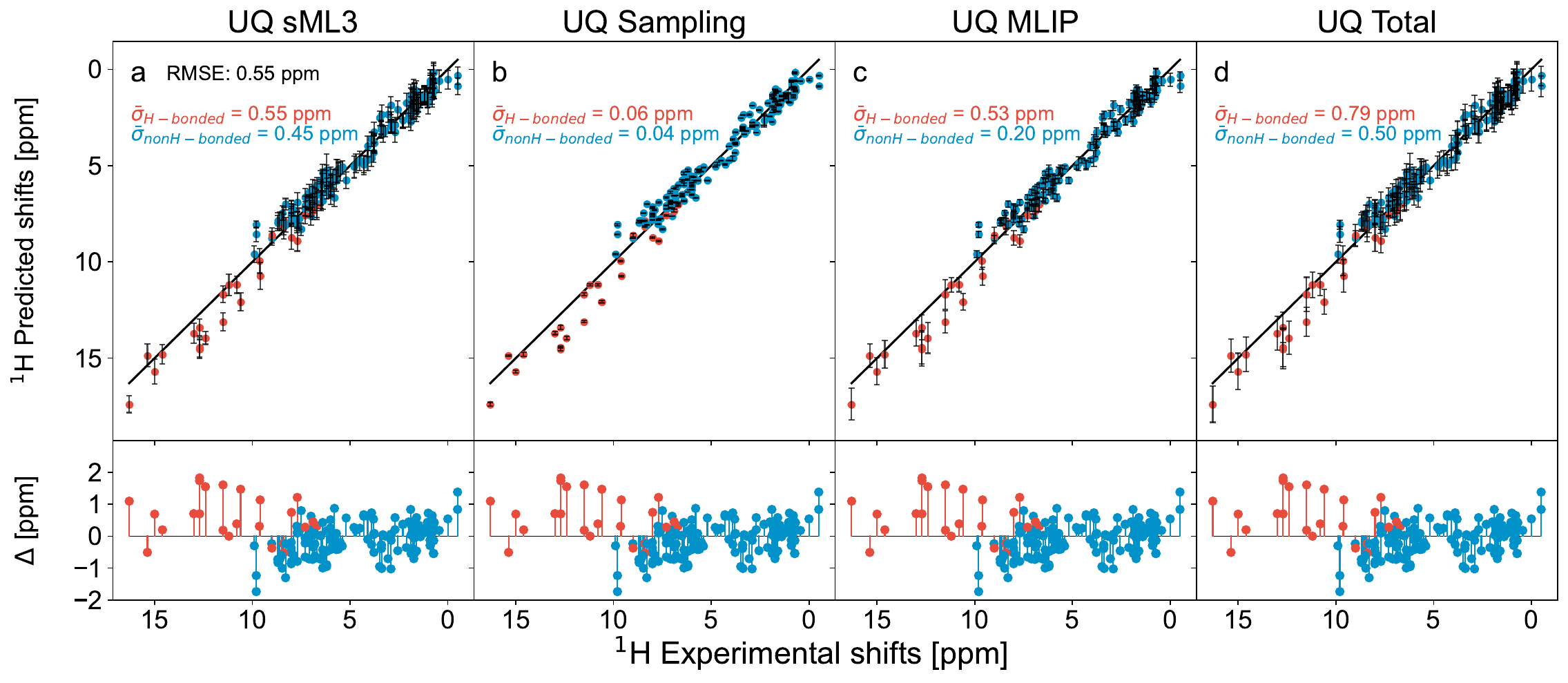}
    \caption{\rev{Comparison of experimental and predicted 1H chemical shifts computed with ShiftML3 using QNC-NMR approach together with the simulation uncertainties on the proton benchmark set, resolved for hydrogen- and non-hydrogen bonded protons. (a) ShiftML3 induced uncertainties on the predictions; (b) the finite time sampling uncertainties; (c) the \petpotential{} induced uncertainties; and (d) the total uncertainties assuming uncorrelated uncertainties. }}
    \label{fig:uq_qnc_nmr_propagated}
\end{figure}

\rev{The use of ML surrogates will inadvertently introduce additional simulation uncertainties, given that the predictions of ML models will be affected by some degree of uncertainty, stemming from either the inability of the model to exactly capture all interactions (as induced by a finite size cutoff) or insuficient training set coverage. On the other hand, ML models also help to mitigate other error sources of uncertainties in the simulation pipeline, namely the simulation uncertainty induced by finite sampling durations, given that it is possible to greatly extend simulation durations using faster ML surrogates.
Separating and quantifying the various sources of uncertainties in the QNC-NMR protocol therefore is instructive, both to understand remaining devations of the simulated shift averages from experiment and also to identify further avenues of improvement for the QNC-NMR protocol
In Figure~\ref{fig:uq_qnc_nmr_propagated} we plot the proton chemicals shift benchmark shifts with simulation uncertainties resolved according to the ShiftML3 shielding surrogate ML model uncertainties, the \petpotential{} propagated uncertainties (via CEA reweighting) and finite sampling uncertainties, and finally the total simulation uncertainties assuming that the individual simulation ingredient uncertainties are uncorrelated. Resolving the uncertainties according to H-bonded and non H-bonded protons indicates, moderate average total uncertainties of (0.5-0.79 ppm) in line with the total RMSE observed of 0.55 ppm. The H-bonded total simulation uncertainties are elevated (0.79 ppm) which can be mostly attributed to an elevated contribution stemming from the \petpotential{} uncertainty contribution. This is in line with the observation in the main-text that the free energy landscape of proton transfer is more sensitive to MLIP errors.

Note however, that these uncertainty estimates ultimately reflect only the uncertainties with respect to a hypothetical ab-initio molecular dynamics simulation that was run until convergence, but the uncertainty quantifiers do not capture the error induced by the DFT-xc functional against reality.  }

\subsection{\rev{Comparing \petpotential{} and ab-initio geometry relaxations }}

\begin{figure}[h!]
    \centering
    \includegraphics[width=0.6\linewidth]{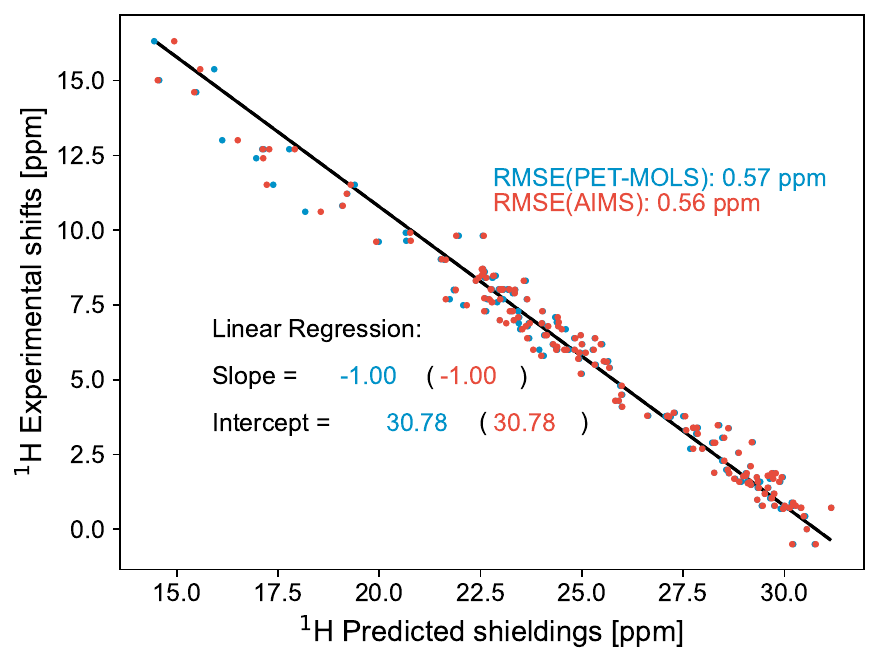}
    \caption{\rev{$^1$H experimental chemical shifts as a function of the ShiftML3 predicted shieldings on \petpotential{} and ab-initio relaxed geometries of 12 structures from the proton benchmark presented in the main text. Note that penicillin salt was removed from the analysis due to ab initio convergence issues. Linear regression parameters keeping the slope fixed to -1 and RMSE are presented for structures relaxed using PET-MOLS and AIMS in blue and red, respectively.}}
    \label{fig:mlip_vs_pet_mols}
\end{figure}

\rev{We compare the quality of \petpotential{} relaxed geometries against full ab-initio FHI-AIMS geometry relaxations. FHI-AIMS calculations are performed with identical parameters as were used to generate the \petpotential{} training database. We exclude penicillin salt from the analysis due to DFT convergence issues. In Figure~\ref{fig:mlip_vs_pet_mols} we plot the experimental chemcial shifts mapped as a function of the ShiftML3 predicted shieldings of both ab-inito and \petpotential{} geometries. The prediction accuracies are in good agreement (0.56 vs 0.57 ppm). We also compute root-mean-squared-deviations (RMSD) between both sets of geometries and find average RMSDs of 0.04~\AA between ab-initio and \petpotential{} geometries. }

\newpage
\subsection{Shielding distribution in succinic acid}

\begin{figure}[h!]
    \centering
    \includegraphics[width=\linewidth]{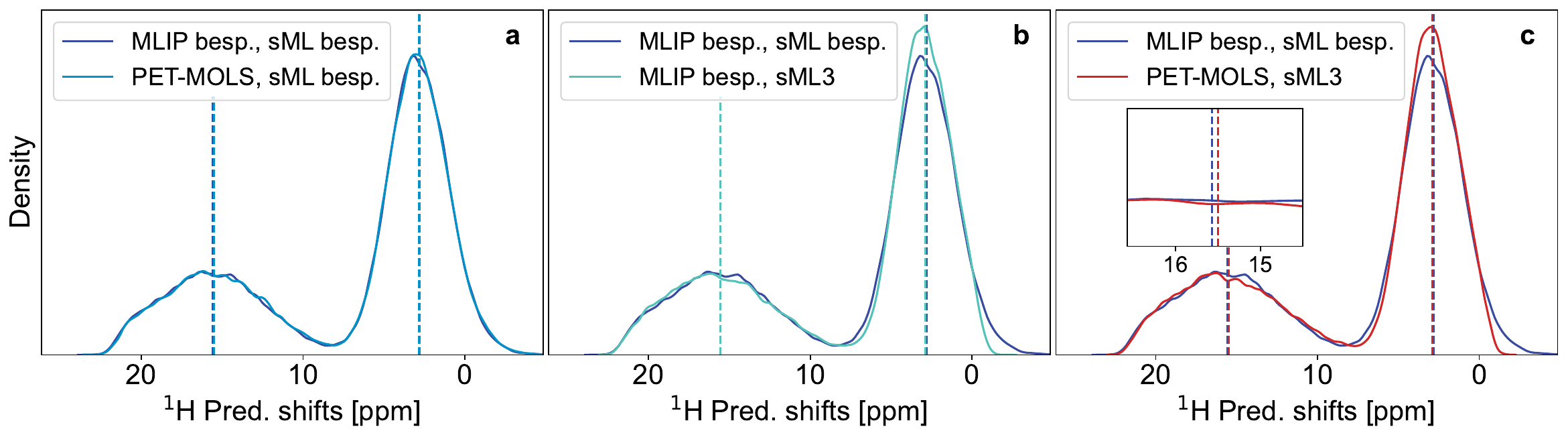}
    \caption{(a) $^1$H chemical shielding distributions computed by bespoke shielding models, but generateing the ensemble of structures using either a bespoke MLIP or \petpotential{}, (b) using the bespoke MLIP to generate an ensemble of structures but computing shieldings with either bespoke or general ShiftML3 chemical shielding model and finally (c), comparing distributions using either entirely bespoke models for structure generation and shielding computations, or the general ShiftML3 and \petpotential{} models, effectively comparing the general QNC-NMR protocol with a fully bespoke one for succinic acid. Dashed lines correspond to averaged shift values of acidic and aliphatic protons in succinic acid.}
    \label{fig:succinic_acid_distributions}
\end{figure}

\begin{figure}[h!]
    \centering
    \includegraphics[width=\linewidth]{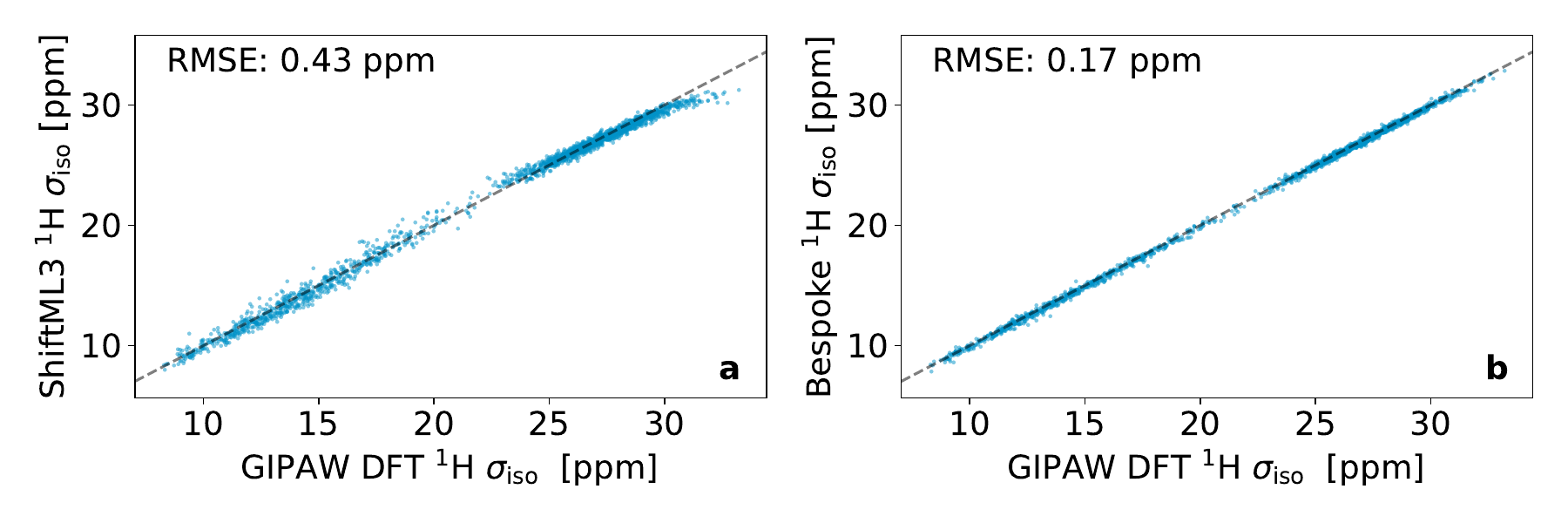}
    \caption{Parity plots of $^1$H chemical shielding predictions of ShiftML3 (a) and from a bespoke machine learning model (b) against reference GIPAW DFT shielding computations.}
    \label{fig:parity_plots_chemical_shieldings_succinic}
\end{figure}

\begin{figure*}[h!]
    \centering
    \includegraphics[width=\columnwidth]{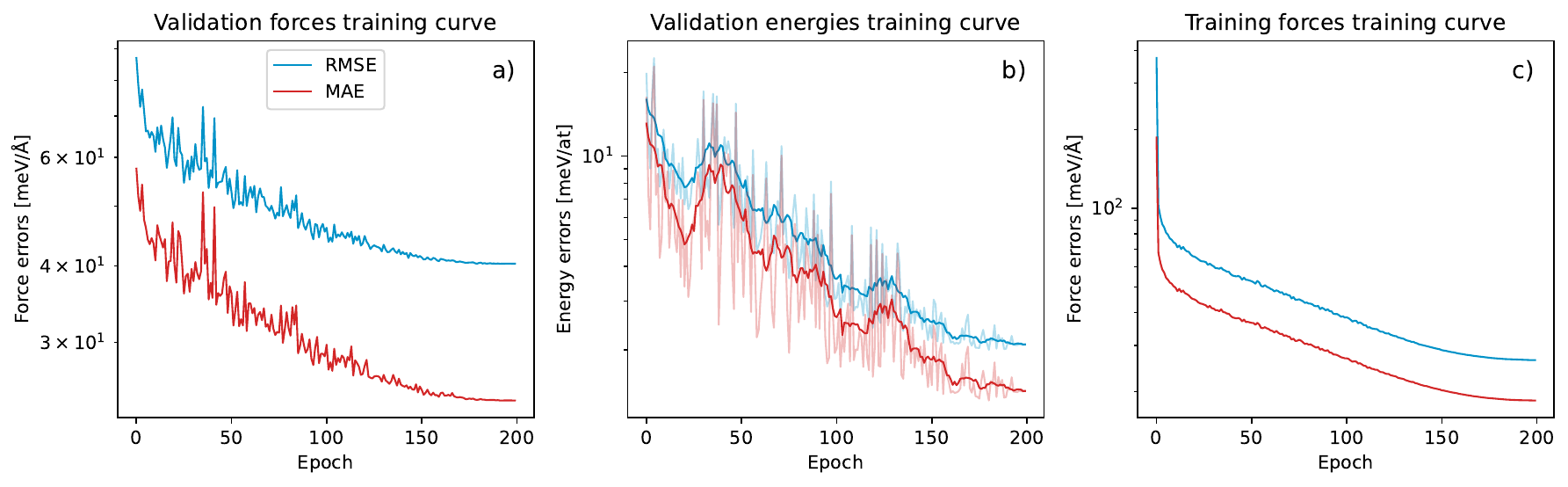}
    \caption{ a) Validation forces and (b) energy error metrics (MAE and RMSE) training curves against epoch number as well as (c) force training set evaluation error, training curve.}
    \label{fig:training_curve}
\end{figure*}

We evaluate our QNC-NMR protocol (\petpotential{} and ShiftML3) against a computational model to assert that the introduction of surrogate models that enable QNC-NMR in the first place does not introduce biases in the computation of shieldings and ensemble averages of shieldings, with respect to the reference electronic structure methods employed for shielding computations and the potential energy surface. Unfortunately, even for the smallest of systems, computing thermodynamic averages of observables using ab initio molecular dynamics at the PBE0+MBD level of theory as well as computing chemical shieldings from the structural ensemble generated by PIMD, becomes prohibitively expensive. Instead, we  train "bespoke" machine learning models that have improved prediction accuracies for both potential energy predictions and shieldings, by training on datasets curated for the purpose of generating highly accurate surrogate models for one specific system, effectively setting up a bespoke QNC-NMR workflow for one specific system. We choose to study the $^1$H chemical shielding distributions in $\alpha$-succinic acid as the reference system. Engel and coworkers, have studied the shielding distributions of acidic protons in succinic acid by means of bespoke MLIPs and chemical shielding models.\cite{enge+21jpcl} The reference DFT computations for their MLIPs were used employing identical PBE0+MBD parameters (FHI-AIMS, medium basis) enabling us to directly compare the potential energy prediction accuracies of \petpotential{} against a densely sampled reference set of DFT computations of succinic acid structures. For chemical shielding predictions we use a linear-SOAP model that we refitted on an additional set of GIPAW-PBE reference computations that were introduced by Engel et al. to build a bespoke chemical shielding model for succinic acid, with comparable $^1$H prediction accuracies as the reference kernel model, presented in the original work of Engel. \cite{mazitovPETMADLightweightUniversal2025a,enge+21jpcl} We construct a bespoke MLIP for succinic acid, by Low-Rank Adaptation (LORA) finetuning \cite{huLoRALowRankAdaptation2021} \petpotential{} on a dataset of 1,800 training configurations of succinic acid computed at the PBE0+MBD level of theory. We refer to this model as LORA-\petpotential{}.~Prediction accuracies of bespoke shielding models and MLIP as well as \petpotential{} are listed in Table~\ref{tab:E_F_accuracy_bespoke_succinic}. Prediction accuracies of the bespoke chemical shielding model and ShiftML3 are listed in Table~\ref{tab:sigma_accuracy_bespoke_succinic}. We show parity plots of ShiftML3 predicted chemical shieldings on a test set of 177 succinic acid configurations and predictions from the bespoke chemical shielding model against GIPAW reference calculations in Fig.~\ref{fig:parity_plots_chemical_shieldings_succinic}.
Dataset sized for both training a chemical shielding and MLIP model are summarized in Table~\ref{tab:succinic_bespoke}. The datasets  were obtained by generating a large number of thermally distorted configurations of $\alpha$ and $\beta$ polymorphs of crystalline succinic acid. More details about the training set generation can be found in the original publication Ref.~\cite{enge+21jpcl}.
\rev{As expected, we find that the bespoke shielding model is more accurate against reference DFT calculations than ShiftML3. Remarkably \petpotential{} achieves energy and force evaluation metrics on hold-out succinic acid test data, that are comparable to those of the finetuned LORA model, highlighting the transferability of \petpotential{}. Both \petpotential{} and the LORA finetuned potential make more accurate force and potential energy predictions than the originally published  MLIPs from Kapil et al.\cite{kapilCompleteDescriptionThermodynamic2022a} (Energy RMSE 3.6 meV/atom and Force RMSE $\approx$ 45 meV/\AA{}).  }
In Fig~\ref{fig:succinic_acid_distributions} we compare $^1$H chemical shielding distributions computed by either bespoke chemical shielding models or ShiftML3, furthermore we compare distributions of chemical shieldings obtained from PIMD simulations employing the finetuned MLIP or the general \petpotential{}. We perform NVT-PIMD molecular dynamics for 250~ps at 300~K with a timestep of 0.5 fs, using a PILE-L langevin thermostat for thermostating with a coupling time of 100~fs. We find computed thermodynamic averages of acidic proton shieldings between ShiftML3/\petpotential{} simulations and bespoke models to be in excellent agreement and only differ by 0.07~ppm, justifying the use of universal MLIPs and chemical shielding models for the computation of thermodynamic averages of chemical shieldings.

\begin{table}[h!]
    \centering
    \caption{Summary of Succinic acid datasets used to train bespoke chemical shielding models and MLIPs. We list the number of succinic acid configurations, for which shieldings (GIPAW-PBE) or energies and forces (PBE0+MBD) have been computed, in training set $n_{\text{train}}$, validation set $n_{\text{val}}$ and test set $n_{\text{test}}$.}
    \label{tab:succinic_bespoke}
    \resizebox{0.35\textwidth}{!}
    {\begin{tabular}{lccc}
        \toprule
        Ref. Theory & $n_{\text{train}}$& $n_{\text{val}}$ & $n_{\text{test}}$ \\
        \midrule
        PBE0+MBD     & 1800 & 200 & 200\\
        GIPAW-PBE & 1450 & 93 & 177 \\
        \bottomrule
    \end{tabular}}
\end{table}

\begin{table}[h!]
    \centering
    \caption{Evaluation metrics of the LoRA-\petpotential{}finetuned model and \petpotential{} on a testset of 200 distorted succinic acid configurations computed at the PBE0+MBD reference level. }
    \label{tab:E_F_accuracy_bespoke_succinic}
    \resizebox{\textwidth}{!}{\begin{tabular}{lcccc}
        \toprule
        Models & E MAE/at. [meV/at.] & E RMSE/at. [meV/at.] & F MAE [meV/Å] & F RMSE [meV/Å] \\
        \midrule
        \petpotential{}      & \rev{0.99} & \rev{1.11} & \rev{20.47} & \rev{27.80} \\
        LoRA-\petpotential{} & 0.65 & 0.80 & 27.95 & 37.38 \\
        \bottomrule
    \end{tabular}}
\end{table}

\begin{table}[ht]
    \centering
    \caption{Evaluation metrics of $^1$H isotropic chemical shieldings predicted by ShiftML3 and a bespoke machine learning model on a testset of 177 distorted succinic acid configurations computed at the GIPAW-PBE level of theory.}
    \label{tab:sigma_accuracy_bespoke_succinic}
    \resizebox{0.5\textwidth}{!}
    {\begin{tabular}{lcc}
        \toprule
        Models & $^1$H $\sigma_{\text{iso}}$ MAE [ppm] & $^1$H $\sigma_{\text{iso}}$ RMSE [ppm] \\
        \midrule
        ShiftML3     & 0.32 & 0.43 \\
        Bespoke & 0.11 & 0.17 \\
        \bottomrule
    \end{tabular}}
\end{table}

\pagebreak
\pagebreak
\section{\petpotential{} Training curve}

\pagebreak
\section{\petpotential{} uncertainty quantifaction. }

\begin{figure}[h!]
    \centering
    \includegraphics[width=0.6\linewidth]{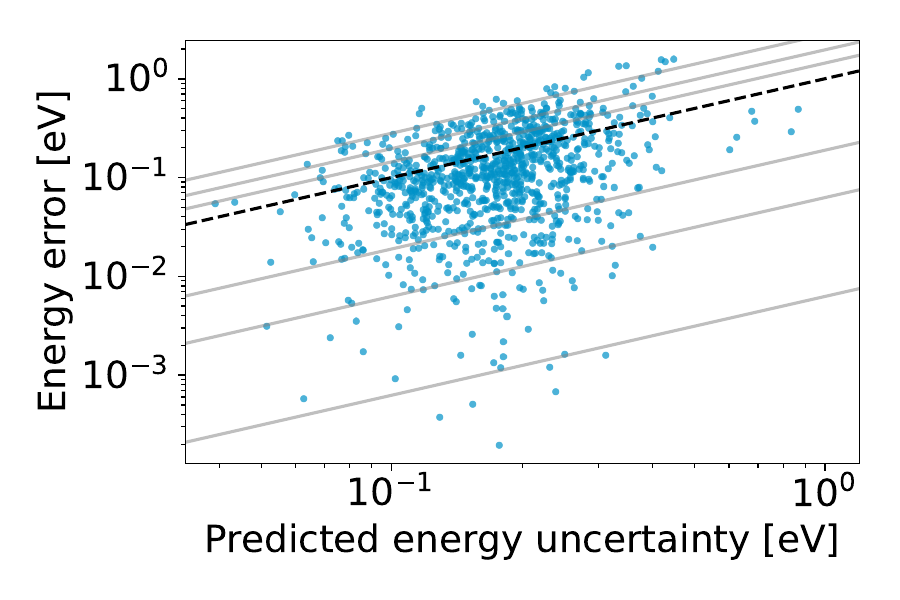}
    \caption{ \petpotential{} predicted logarithmic potential energy uncertainties against logarithmic prediction errors of \petpotential{}. The grey lines represent quantiles of the expected error distribution at each predicted uncertainty level and serve as a visual reference. For instance, globally across the test set predictions, well-calibrated uncertainty estimates should place 90\% of samples between the two outermost lines, which correspond to the 95th and 5th quantiles of the log-folded normal distribution. }
    \label{fig:uq_z_Plot_V}
\end{figure}

We equip \petpotential{} with an uncertainty estimator of its potential energy predictions. We use the last-layer prediction rigidity approach, which is a last layer Laplace approximation to the neural network weight posterior \cite{bigiPredictionRigidityFormalism2024}. We then sample an ensemble of last layer weights, from the approximated last layer posterior. Last layer ensembling has been proven effective for the uncertainty quantification of atomistic simulation, particularly for cases in which the neural network predictions of the potential energy $V_{\text{NN}}(A)$ for a structure, needs to be propagated to an arbitrary function $f(V_{\text{NN}}(A))$ (for example the thermodynamic averages of chemical shieldings).\cite{kellnerUncertaintyQuantificationDirect2024a} We find that a double logarithmic plot, in which we plot predicted potential energy uncertainties against unsigned prediction errors on some hold out test set, provides a valuable visual aid to asses the quality of uncertainty estimators. In Figure~\ref{fig:uq_z_Plot_V} we present such a plot of the \petpotential{} predicted uncertainties (the standard deviation of the \petpotential{} committee energy predictions) against the unsigned prediction errors. 
For more information how to interpret these plots, please also consult References \cite{kellnerUncertaintyQuantificationDirect2024a, bigiPredictionRigidityFormalism2024}.

    \label{fig:15N_shieldings_UQ}


\section{Learning from experiments: Additional tests}

\begin{figure}[h!]
\centering

\begin{minipage}{0.48\textwidth}
  \centering
      \includegraphics[width=\linewidth]{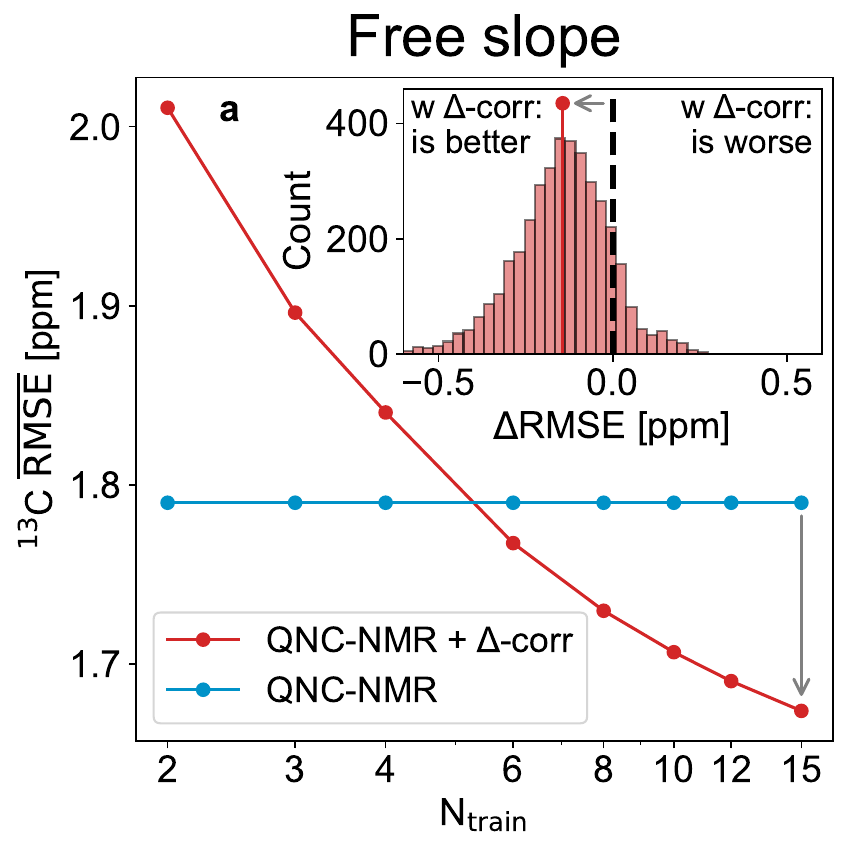}
  \caption*{(a) Learning curve free slope}
\end{minipage}\hfill
\begin{minipage}{0.48\textwidth}
  \centering
  \includegraphics[width=\linewidth]{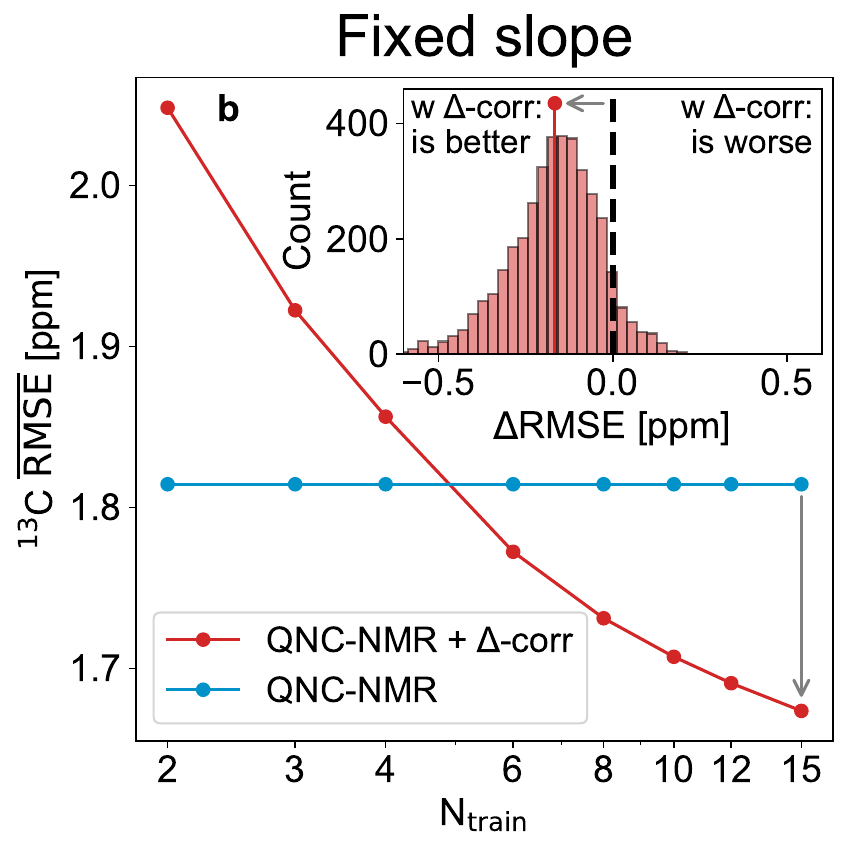}
  \caption*{(b) Learning curve fixed slope}
\end{minipage}

\vspace{0.5em}

\begin{minipage}{0.48\textwidth}
  \centering
  \includegraphics[width=\linewidth]{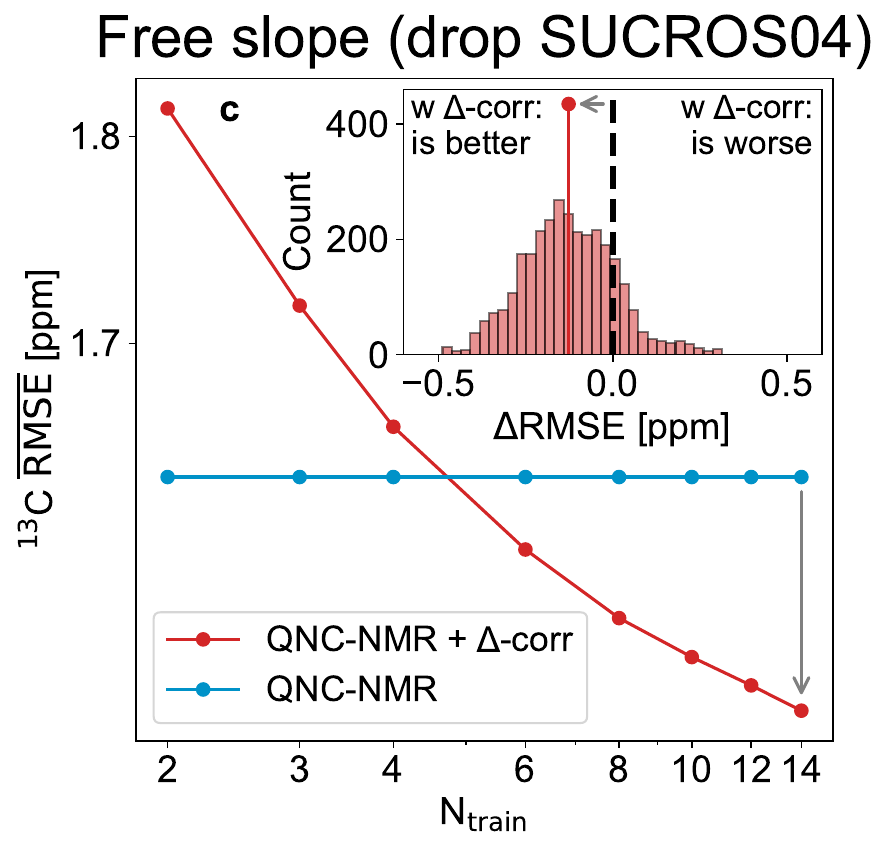}
  \caption*{(c) Learning curve free sloped, after removing SUCROS04.}
\end{minipage}\hfill
\begin{minipage}{0.48\textwidth}
  \centering
  \includegraphics[width=\linewidth]{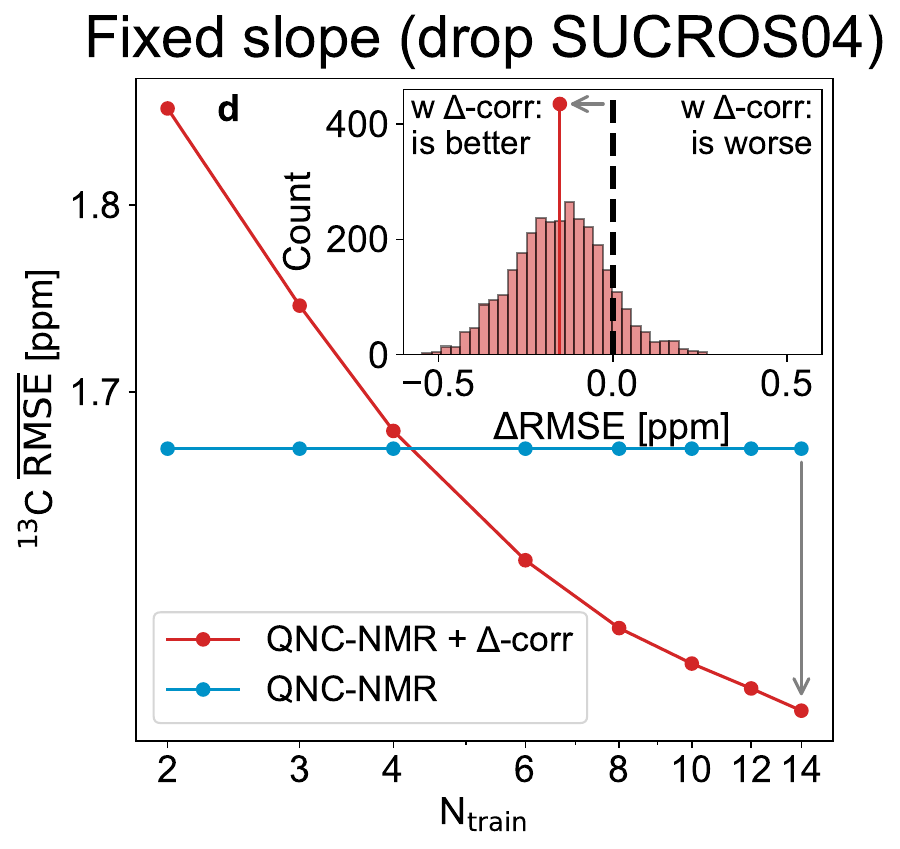}
  \caption*{(d) Learning curve fixed sloped, after removing SUCROS04.}
\end{minipage}

\caption{Learning curves of corrected QNC-NMR $^{13}$C chemical shift predictions from experiments, probing the influence of fixed/free slope and removing SUCROS04 from the training and testing data. The inset shows a histogram of the change in prediction accuracy, $\Delta \mathrm{RMSE}$, between uncorrected and corrected predictions, computed over all $\binom{19}{4}$  splits  of the 19 benchmark structures (or $\binom{18}{4}$ when SUCROS04 is removed) with a fixed test-set size of four structures. Splits to the left of the dashed line correspond to cases where the experimentally corrected QNC-NMR shielding predictions are more accurate on the test set, whereas splits to the right indicate cases where the correction reduces the prediction accuracy.}
\label{fig:comaprison_slopes}
\end{figure}

In this section we present additional experiments probing the stability of the training exercise correcting QNC-NMR with experimental data. In Fig.~\ref{fig:comaprison_slopes} we present learning curves applying the ELF framework to correct $^{13}$C chemical shift predictions from experimental data. The experimental shifts are listed in Table~\ref{tab:table_13C}, and are identical to those used for benchmarking the QNC-NMR protocol.
Corrected last layer weights $w_{\text{corr}}$ are determined by  first converting QNC-NMR predicted shieldings to shifts, employing regression parameters tabulated in table~\ref{tab:RMSE_bench}. Following this,  the difference between QNC-NMR predicted shifts and experimentally measured values are minimized using the ELF formalism. In figure \ref{fig:comaprison_slopes}, we show the improvement in chemical shift prediction when only a limited number of experimental shifts are included in the training. We also test the robustness of our findings to changing composition of the arguably very small training set. In order to test the robustness, we remove structure SUCROS04 from the pool of training structures, which contains the environment with the largest QNC-NMR prediction errors against experimental measurements (deviating by 9.3 ppm). We find that the conclusion 

\begin{figure}[htbp]
  \centering

  \begin{minipage}{0.48\textwidth}
    \centering
    \includegraphics[width=\linewidth]{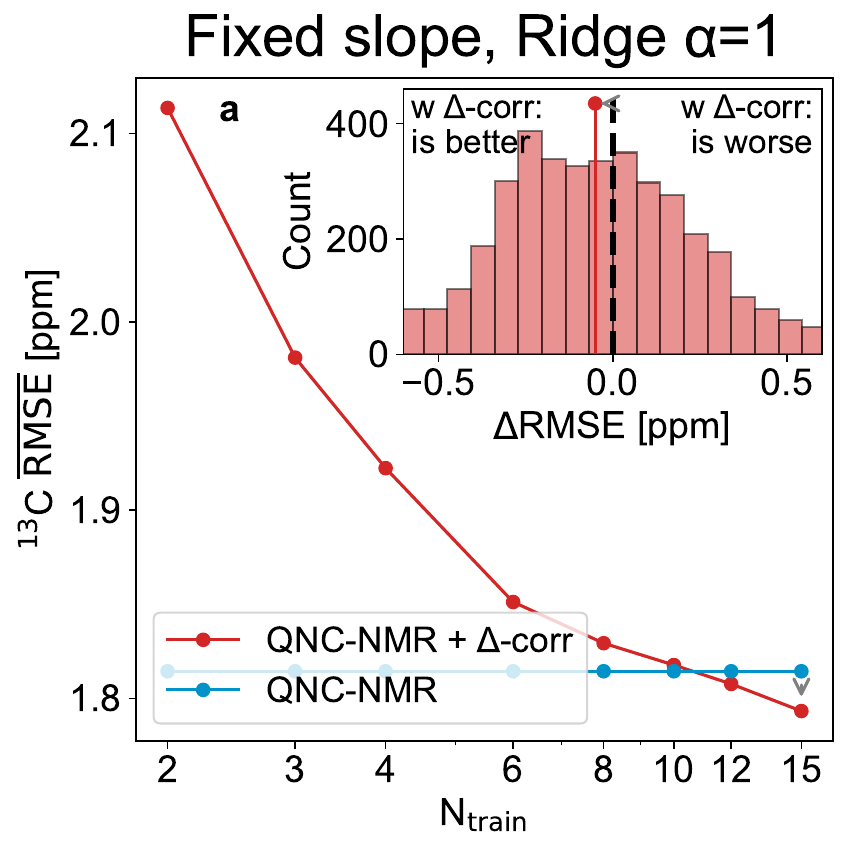}
    \caption*{(a) Learning curve, employing a ridge regressor with regularization strength $\alpha=1$.}
  \end{minipage}\hfill
  \begin{minipage}{0.48\textwidth}
    \centering
    \includegraphics[width=\linewidth]{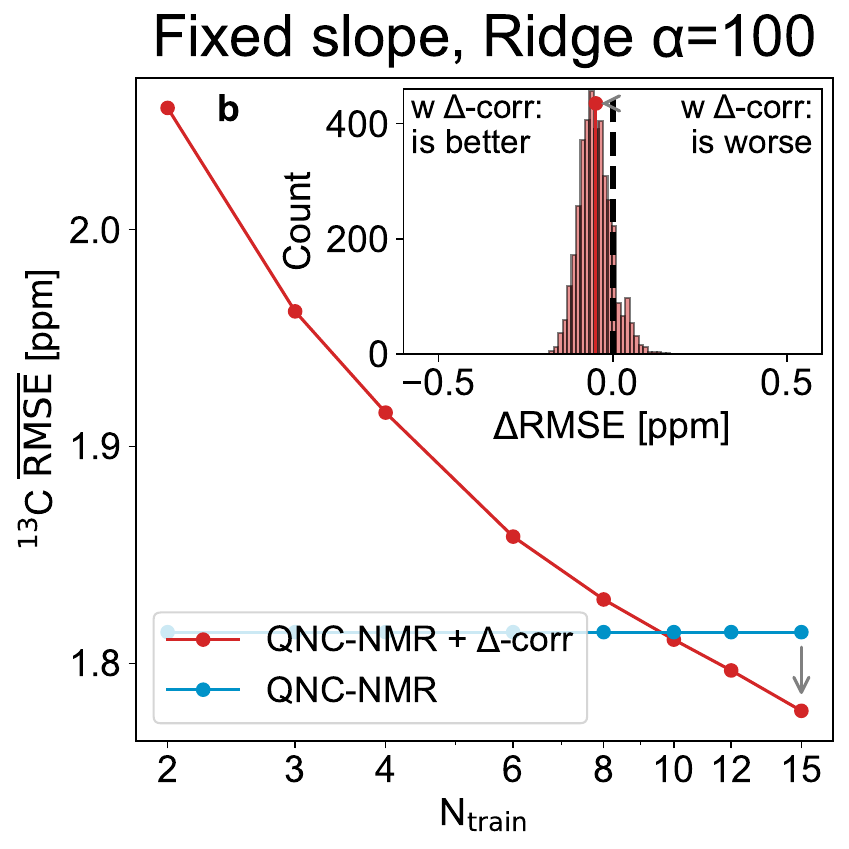}
    \caption*{(b) Learning curve, employing a ridge regressor with regularization strength $\alpha=100$}
  \end{minipage}
    
  \caption{Learning curves of corrected QNC-NMR $^{13}$C chemical shift predictions from experiments, probing the effect of the regularizer strength. The inset shows a histogram of the change in prediction accuracy, $\Delta \mathrm{RMSE}$, between uncorrected and corrected predictions, computed over all $\binom{19}{4}$  splits  of the 19 benchmark structures  with a fixed test-set size of four structures. Splits to the left of the dashed line correspond to cases where the experimentally corrected QNC-NMR shielding predictions are more accurate on the test set, whereas splits to the right indicate cases where the correction reduces the prediction accuracy.}
  \label{fig:comparison_regularizer}
\end{figure}

We also test the effect of varying the regularization strength ($\alpha$) of the ridge regressor employed to find corrected weights. In Figure~\ref{fig:comparison_regularizer} we assess the effect of employing varying regularizer strengths in the finetuning from experiment for $\alpha$ value of 1 and 100.  We find that the overall trend persists that eventually after adding more structures experimentally finetuned QNC-NMR shift predictions become more accurate than uncorrected QNC-NMR predictions.
